\documentclass[prb,showpacs,preprintnumbers,amsfonts,amssymb,amsmath,floats,twocolumn,aps,superscriptaddress]{revtex4-1}

\usepackage{amsmath}
\usepackage{amsthm}
\usepackage{amssymb}
\usepackage{graphicx}%
\usepackage{bm}%
\usepackage{bbold}%
\usepackage{xcolor}%

\usepackage[caption=false]{subfig}
\usepackage[pdftex]{hyperref} %
\definecolor{darkblue}{rgb}{0,0,.5} %
\definecolor{black}{rgb}{0,0,0} %
\hypersetup{
  pdftex=true,
  colorlinks=true, breaklinks=true, linkcolor=darkblue,
  menucolor=darkblue,
  urlcolor=darkblue, citecolor=darkblue }

\def\Tr{\operatorname{Tr}}

\def\mc{\mathcal}
\def\bs{\boldsymbol}

\def\op{}
\def\mat{\bs}
\def\fcl{\widehat}

\def\cre#1#2{#1^\dagger_{#2}}%
\def\ann#1#2{#1^{\vphantom{\dagger}}_{#2}}%

\def\cc#1{\cre{c}{#1}}%
\def\ac#1{\ann{c}{#1}}%
\def\n#1{n_{#1}}%
\def\ra{\rightarrow}%
\def\ua{\uparrow}%
\def\da{\downarrow}%

\def\com[#1,#2]{\left[#1,#2\right]}
\def\contcom[#1,#2]{\left[#1\stackrel{\circ}{,}#2\right]}

\def\ev<#1>{\left<#1\right>}

\def\var{\Delta_{1{\rm D}}}

\begin{document}

\title{Nonequilibrium self-energy functional approach to the dynamical
  Mott transition}
  
\author{Felix Hofmann}%
\email{fhofmann@physik.uni-hamburg.de}%
\affiliation{I. Institute of Theoretical Physics, University of
  Hamburg, 20355 Hamburg, Germany}%
\author{Martin Eckstein}%
\affiliation{Max Planck Institute for the Structure and Dynamics of
  Matter, 22761 Hamburg, Germany}%
\affiliation{University of Hamburg-CFEL, 22761 Hamburg, Germany}%
\author{Michael Potthoff}%
\affiliation{I. Institute of Theoretical Physics, University of
  Hamburg, 20355 Hamburg, Germany}

\begin{abstract}
  The real-time dynamics of the Fermi-Hubbard model, driven out of
  equilibrium by quenching or ramping the interaction parameter, is
  studied within the framework of the nonequilibrium self-energy
  functional theory.  A dynamical impurity approximation with a single
  auxiliary bath site is considered as a reference system and the
  time-dependent hybridization is optimized as prescribed by
  the variational principle. The dynamical two-site approximation
  turns out to be useful to study the real-time dynamics on short and
  intermediate time scales. Depending on the strength of the
  interaction in the final state, two qualitatively different response
  regimes are observed. For both weak and strong couplings, qualitative
  agreement with previous results of nonequilibrium dynamical
  mean-field theory is found. The two regimes are sharply separated by
  a critical point at which the low-energy bath degree of freedom
  decouples in the course of time.  We trace the dependence of the
  critical interaction of the dynamical Mott transition on the
  duration of the interaction ramp from sudden quenches to adiabatic
  dynamics, and therewith link the dynamical to the equilibrium Mott
  transition.
\end{abstract}


\maketitle

\section{Introduction}
\label{sec:introduction}

Systems of strongly correlated electrons on a lattice exhibit diverse
emergent phenomena such as insulating behavior caused by strong local
Coulomb repulsion.  The Mott insulator is believed to hold the key for
an understanding of the complex physics that is characteristic for
several transition metals and compounds and has been at the focus of a
vast number of studies in recent decades.\citep{imada1998} The
preparation and the study of a Mott-insulating state by experiments
done with ultracold fermionic atoms \citep{joerdens2008,schneider2008}
opens an exciting new perspective on the underlying many-body problem
and on the related idealized many-body models, such as the single-band
Fermi-Hubbard model.\citep{giamarchi2004} The new aspect in those
experiments is the high degree of control over the microscopic model
parameters which can be used to steer the system between different
phases in the equilibrium phase diagram but also to initiate and to
manipulate nonequilibrium processes. \citep{bloch2008,lewenstein2012}
The last point in particular has attracted some interest recently, and
time-dependent experiments with ultracold atoms in optical lattices
\citep{greiner2002,strohmaier2010} can complement the investigation of
condensed-matter dynamics on femtosecond time scales with ultrafast
pump-probe experiments.  \citep{iwai2003,perfetti2006,wall2011}

The study and the understanding of strongly interacting
lattice-fermion systems far from equilibrium requires a critical
examination of standard concepts of quantum statistics regarding,
e.g., the thermalization of isolated quantum
systems,\citep{deutsch1991,srednicki1994,rigol2008} and it can bring
about entirely new concepts such as dynamical phase
transitions. \citep{dziarmaga2010,polkovnikov2011} Apart from such
fundamental theoretical questions, a further development and
application of numerical methods is highly needed to study relevant
problems such as the correlation-driven metal-insulator transition.

The Mott transition in the single-band Hubbard model is the
paradigmatic field of application for the dynamical mean-field theory
(DMFT),\citep{metzner1989,georges1996} which may be characterized as
an internally consistent and nonperturbative mean-field approach
controlled by the limit of infinite spatial dimensions. Although the
feedback of nonlocal magnetic correlations is neglected, the DMFT
phase diagram of the (paramagnetic) Mott transition
\citep{georges1996} represents an instructive example of a phase
transition which must be described by nonperturbative means.
This has triggered the study of the real-time dynamics at the Mott
transition using the extension of the standard DMFT to the
nonequilibrium case.\citep{schmidt2002,freericks2006,aoki2013} The
simplest protocol to initiate the dynamics is a sudden quench of the
Hubbard-$U$, from $U_{\rm ini}=0$ to different final values $U_{\rm
  fin}$.  For weak $U_{\rm fin}$, it has been found that
thermalization is delayed and that the system gets trapped in a
so-called prethermal metastable
state.\citep{moeckel2008,eckstein2009b} On the other hand, for strong
$U_{\rm fin}$, a fast relaxation is again impeded by
collapse-and-revival oscillations that are reminiscent of the dynamics
in the atomic limit.  Both regimes are separated by a sharp transition
for a certain final interaction strength $U_{\rm fin} = U_{\rm c}^{\rm
  dyn}$, at which a fast relaxation to thermal equilibrium takes
place.  This picture of the ``dynamical Mott transition'' has emerged
by applying the nonequilibrium DMFT to the Hubbard model on an
infinite-dimensional Bethe lattice, \citep{eckstein2009b} and has been
corroborated by different subsequent studies and using different
methods.\citep{schiro2010b,schiro2011,sandri2012,hamerla2013,hamerla2014}

Within the DMFT the original lattice-fermion system is mapped onto an
impurity model where a single correlated site is embedded in a
noninteracting dynamical mean field (``bath'') which must be
determined self-consistently.  This effective impurity model poses a
demanding many-body problem, particularly for a general nonequilibrium
situation.  While continuous-time quantum Monte-Carlo methods
represent highly efficient ``solvers'' for the equilibrium case, only
short propagation times can be accessed due to a severe sign (or
phase) problem showing up in the real-time
domain. \citep{werner2009,gull2011} Perturbative approaches are much
more successful and have been used extensively in the strong-
\citep{eckstein2010} and weak-coupling regime
\citep{eckstein2011b,tsuji2013b} but are clearly of limited use to
address the Mott transition which takes place at intermediate coupling
strengths. As concerns the single-band Hubbard model, an
exact-diagonalization (ED) solver is an efficient method for the
equilibrium case at zero temperature,\citep{caffarel1994} though the
determination of the one-particle parameters of the impurity
Hamiltonian is essentially done in an {\em ad hoc} way.  For a
nonequilibrium problem, the necessary Hamiltonian representation of
the effective mean field poses an even more severe complication which
has so far been solved on a short time scale only \citep{gramsch2013}
since in the course of time more and more bath degrees of freedom have
to be coupled to the impurity.  Still this has allowed ED approaches
to operate, such as Krylov-space methods, \citep{gramsch2013} the
multiconfiguration time-dependent Hartree method \citep{balzer2015}
or density-matrix renormalization techniques based on matrix-product
states.\citep{wolf2014}

The (nonequilibrium) self-energy functional theory
(SFT),\citep{potthoff2003,hofmann2013} offers a different route for
the application of ED methods.  Here, a reference impurity model with
a given finite (small) number of bath degrees of freedom is
considered.  Instead of imposing the DMFT self-consistency condition,
the time-dependent parameters of the reference system are fixed by
applying a general variational principle, stating that the grand
potential should be stationary as when expressed as a functional of
the (nonequilibrium) self-energy.  While the DMFT is recovered by
choosing a reference system with a continuum of bath sites, any
reference system with a finite bath generates a consistent dynamical
impurity approximation (DIA) or, when choosing, in the spirit of
cluster-DMFT approaches, a finite cluster of correlated sites to
better account for short-range correlations, a variational cluster
approximation (VCA). For an overview we refer to
Refs.~\onlinecite{potthoff2012,potthoff2014}.

This idea has successfully been employed for the study of the
(equilibrium) Mott transition.  Remarkably, the DIA with only a single
additional bath site has turned out to recover the DMFT picture of the
Mott transition in a qualitatively correct way,
\citep{potthoff2003b,pozgajcic2004,koga2005,inaba2005b,eckstein2007}
and with a few more bath degrees of freedom \citep{pozgajcic2004} the
agreement is even quantitative.

The present paper reports on results obtained by applying the
nonequilibrium extension of the dynamical impurity approximation
to study the real-time dynamics after quenches and ramps of the
Hubbard interaction.  For the nonequilibrium case, the fundamental
concepts of self-energy functional theory are somewhat different and
require a completely new strategy for the numerical
evaluation.\citep{hofmann2013} Different practical issues of the
implementation have been discussed recently in
Ref.~\onlinecite{hofmann2015}.  For the first application of the
nonequilibrium DIA we
restrict ourselves to a two-site reference system.  As will be
discussed, this is indeed sufficient in many respects to cover the
essentials of the dynamical Mott transition as compared to previous
work.\citep{eckstein2009b,schiro2010b} We also discuss the
significance of the method in view of recently proposed
Hamiltonian-based impurity
solvers.\citep{gramsch2013,balzer2015,wolf2014}

The paper is organized as follows: In Sec.~\ref{sec:model-methods}
we briefly introduce the nonequilibrium many-body problem, review the
self-energy functional approach, and discuss the cornerstones of its
numerical implementation.  The application to study the Mott
transition for both the equilibrium and the nonequilibrium case is
presented in Secs.~\ref{sec:equil-mott-trans} and \ref{sec:dynamit},
respectively.  Section~\ref{sec:conclusion} provides a summary.

\section{Model and Methods}
\label{sec:model-methods}
Using standard notations, the Hamiltonian of the Fermi-Hubbard model
at half-filling reads
\begin{equation}
  \label{eq:HubbardHamiltonian}
  \op H(t) = -T\sum_{\langle ij \rangle,\sigma}\cc{i\sigma}\ac{j\sigma} +
  U(t) \sum_{i} \left(\n{i\ua}-\frac{1}{2}\right)\left(\n{i\da}-\frac{1}{2}\right) \,.
\end{equation}
Here, $c^{(\dagger)}_{i\sigma}$ (creates) annihilates a fermion at
site $i$ and with spin projection $\sigma=\ua,\da$, and the number
operator is given by $\n{i\sigma} =
\cc{i\sigma}\ac{i\sigma}$. Fermions can tunnel between neighboring
sites $\langle ij\rangle$ with the hopping amplitude $T$.  Two
fermions on the same site are repelled by the local Coulomb
interaction $U$. Nontrivial real-time dynamics can be stimulated by
controlling the explicit time dependencies of the model parameters.
Here, we investigate both sudden quenches and ramps of different
duration of the interaction parameter $U(t)$.

\begin{figure}[t]
  \centering
  \includegraphics[width=.8\columnwidth]{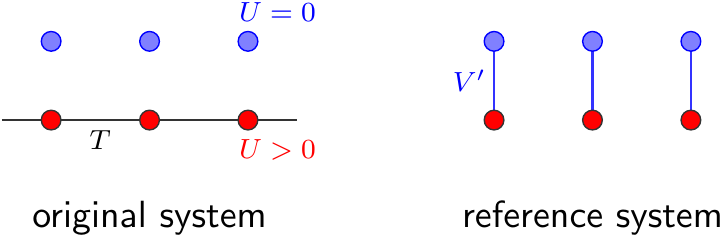}%
  \caption{Illustration of the original (left) and the reference
    system (right). Black solid lines indicate nearest-neighbor
    hopping between correlated sites (red filled dots). The original
    system is the Hubbard model with additional uncorrelated sites
    (blue filled dots) which, however, are decoupled and merely ensure
    that both Hamiltonians operate in the same Hilbert space.  In the
    reference system (right) these ``bath sites'' are locally coupled
    to the otherwise disconnected correlated sites via a hybridization
    parameter $V'(t)$, indicated by blue solid lines.}
  \label{fig:orgref}
\end{figure}

Calculations have been performed within the two-site dynamical
impurity approximation (DIA), which provides a \emph{local}
approximation to the self-energy. It relates the full lattice problem
to a small reference system consisting of a single correlated site
($L_c = 1$) sharing the same time-dependent interaction as the
original model but with an additional uncorrelated ``bath'' site ($L_b
= 1$) coupled to it via the time-dependent hybridization $V'(t)$. For
an illustration, see Fig.~\ref{fig:orgref}.  Here and in the
following, primed quantities refer to the reference system.

The parameter $V'(t)$ is determined via a variational principle set up
within the general framework of the nonequilibrium self-energy
functional theory (SFT).\citep{hofmann2013} The latter exploits the
fact that the initial state equilibrium grand potential of the
original system, $\Omega$, at inverse temperature $\beta$ can be
expressed as a functional of the nonequilibrium self-energy. The
self-energy functional is stationary at the \emph{physical}
self-energy $\mat\Sigma$ of the model, i.e., $\delta
\fcl\Omega[\mat\Sigma] = 0$, where it equals the physical
grand potential, namely $\fcl\Omega[\mat\Sigma] = \Omega$. Note that
functionals are indicated by a hat and that $\mat\Sigma$ is defined on
the Keldysh-Matsubara contour $\mc
C$,\citep{wagner1991,leeuwen2006c,rammer2007} i.e., has elements
$\Sigma_{ij,\sigma}(z,z')$ with complex contour times $z,z'$.

Assuming that the problem posed by the reference system can be solved,
the self-energy functional of the original system can be evaluated
exactly for a certain subclass of trial self-energies, namely for the
exact self-energies $\mat\Sigma'$ of the reference system at different
parameters $V'(z)$. We have
\begin{equation}
  \label{eq:SFTfcl}
  \fcl\Omega[\mat\Sigma'] = \Omega' + \frac{1}{\beta}
  \Tr\ln(\mat G_0^{-1}-\mat\Sigma')^{-1} - \frac{1}{\beta} \Tr\ln
  \mat G'\,, 
\end{equation}
where the grand potential $\Omega'$, the self-energy $\mat\Sigma'$, and
the Green's function $\mat G'$ of the reference system are functionals
of $V'(z)$. 
Furthermore, $\beta$ is the inverse temperature of the initial equilibrium state.
One must consider $V'(z)$ with $z$ on the upper and lower
Keldysh branch as independent variables, since otherwise the
functional dependence of $\fcl\Omega$ on all real-time quantities
would disappear. Finally, $\mat G_0$ is the free Green's function
of the original system. 
The trace contains an implicit integration along $\mc
C$, i.e., we defined $\Tr\mat A = \sum_{i\sigma}\int_{\mc C}
dz\,A_{i\sigma,i\sigma}(z,z^+)$, where $z^+$ is infinitesimally later
on $\mc C$. The optimal value $V'_{\rm opt}(t)$ is determined at each
instant of time according to the stationarity principle
\begin{equation}
  \label{eq:EulerEqn}
  \left.\frac{\delta\fcl\Omega[\mat\Sigma']}{\delta
      V'(z)}\right|_{V'(t) = V_{\rm opt}'(t)} = 0\,,
\end{equation}
which is evaluated on the space of \emph{physical} parameters $V'(t)$.
This Euler equation is the central equation of the (two-site)
dynamical impurity approximation, which gives us access to the optimal
local self-energy $\mat\Sigma'_{\rm opt}$ and therewith to the
one-particle Green's function $\mat G^{\rm DIA} \equiv (\mat
G_0^{-1}-\mat\Sigma'_{\rm opt})^{-1}$.
  
Equation (\ref{eq:EulerEqn}) is inherently causal, i.e., optimal
parameters can be determined at some time $t$ without affecting
results at earlier times $t'<t$.  This allows for the implementation
of a time-propagation scheme for $V'_{\rm opt}(t)$. For reasons
discussed in Ref.~\onlinecite{hofmann2013}, it is beneficial to carry
out the functional derivative in Eq.~(\ref{eq:EulerEqn}) analytically,
which turns the variational problem into the problem of finding the
zeros of $\fcl K[V'](t) := -\beta\delta\fcl\Omega[\mat\Sigma']/\delta
V'(z)\big|_{V'(t)}$. The latter, however, proves to be highly
unstable.  \citep{hofmann2015} Fortunately, this problem can be
bypassed by only fixing the initial condition (at $t=t_0$) via $\fcl
K[V'](t_0) = 0$ and for later times $t>t_0$ requiring the respective
time derivative to vanish, i.e., $\partial_t\fcl K[V'](t) = 0$. Thus,
we finally have to solve
\begin{subequations}
  \label{eq:DiffEulerEqn}
  \begin{align}
    \left. \fcl K[V'](t)\right|_{V'=V'_{\rm opt}} &= 0 \,, \qquad
    \text{for } t=
    t_0\,,  \label{eq:DiffEulerEqn_eq} \\
    \left.\partial_t \fcl K[V'](t)\right|_{V'=V'_{\rm opt}} &= 0 \,,
    \qquad \text{for } t> t_0\,. \label{eq:DiffEulerEqn_ne}
  \end{align}
\end{subequations}
For precise details on the SFT framework and its numerical
implementation we refer to Refs.~\onlinecite{hofmann2013,hofmann2015}.

In principle, approximations within the self-energy functional theory
can be constructed such that they respect the macroscopic conservation
laws and the respective continuity equations for particle number, spin,
and energy. Conservation laws for one-particle quantities in fact hold
for any choice of the reference system, but obeying energy
conservation requires a continuum of variational degrees of
freedom.\citep{hofmann2013} Hence, for small reference systems, energy
conservation is weakly violated, but one can expect to gradually
improve on this by adding further variational degrees of freedom. It
is noteworthy, that by providing a continuous bath (i.e., $L_b =
\infty$) one formally recovers the dynamical mean field theory
(DMFT),\citep{georges1996,freericks2006,schmidt2002,aoki2013} (for
$L_c=1$ or its cluster extension for $L_c>1$) which in fact is a fully
conserving approximation.

For our calculations we consider the half-filled Hubbard model on a
one-dimensional lattice of 40 sites with periodic boundary conditions,
which is sufficient to ensure numerically converged
results.\citep{convergence} Choosing a one-dimensional system is
convenient for numerical reasons. It is important to note, however,
that the lattice dimension and geometry enters the DIA only via the
free density of states (DOS).  Moreover, we expect that results
essentially depend on the variance of the DOS
only.\citep{potthoff2003b} For the one-dimensional lattice this is
$\var = \sqrt{2} \,T$.  Energy (and time) units are fixed by setting
$T=1$.  Calculations have been performed for different inverse
temperatures $\beta$, which set the length of the Matsubara branch in
Eqs.~(\ref{eq:SFTfcl})--(\ref{eq:DiffEulerEqn}).  All integrations
over imaginary time $\tau$ have been carried out using accurate
high-order numerical integration schemes with step sizes varying from
$\Delta\tau = 0.05$ for larger $\beta$ to $\Delta\tau = 0.1$ for
smaller $\beta$. For the real-time propagation and integration along
both Keldysh branches, however, we are limited to the trapezoidal
rule.\citep{convergence} Sufficiently converged time propagations are
obtained for time steps $\Delta t = 0.04 \dots 0.05$ for maximum times
up to $t_{\rm max}\leq 25$.


\section{Equilibrium Mott transition}
\label{sec:equil-mott-trans}

Before the real-time dynamics of the Hubbard model can be analyzed
within the two-site DIA, a proper initial state has to prepared, that
is, the equilibrium variational problem [Eq.~(\ref{eq:EulerEqn}) at
$t=t_0$] has to be solved.  To the best of our knowledge, all previous
SFT studies evaluated the grand potential $\fcl\Omega[\mat\Sigma']$
[cf. Eq.~\ref{eq:SFTfcl}] (and possibly its
derivatives\citep{pozgajcic2004}) directly to search for stationary
points. Here we instead determine the equilibrium solution by
evaluating its derivative analytically and look for the roots of $\fcl
K[V']$ by solving Eq.~(\ref{eq:DiffEulerEqn_eq}). Thus, as a benchmark
and to introduce our method, in this section we reproduce and discuss
the known equilibrium results for a two-site reference system, as
depicted in Fig.~\ref{fig:orgref}.

\begin{figure}[t]
  \centering
  \includegraphics[width=0.9\columnwidth]{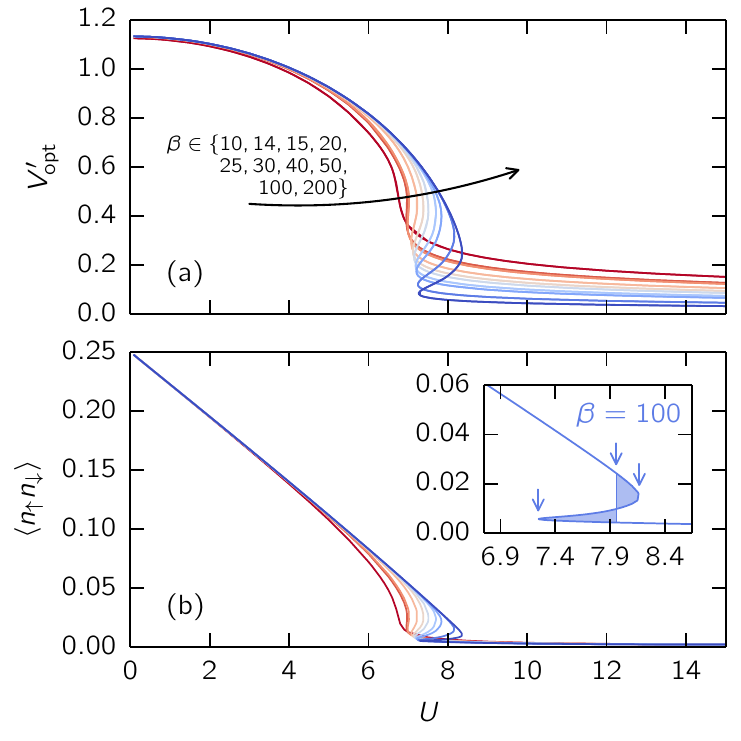}
  \caption{(a) Optimal variational parameter $V'_{\rm opt}$ as a
    function of the interaction for different inverse temperatures
    increasing from $\beta = 10$ (red curve) to $\beta = 200$ (blue
    curve), and (b) the respective double occupancies.  The inset in
    (b) shows the Maxwell construction for $\beta = 100$: The mid
    arrow indicates the value for $U_{\rm c}$, the outer arrows point
    at the spinodal points, which define the region where metallic and
    insulating solutions coexist.}
  \label{fig:eqVdocc}
\end{figure}

Results for the optimal hybridization parameter $V'_{\rm opt}$ are
shown in Fig.~\ref{fig:eqVdocc}(a).  In general, the on-site energies
of the correlated and of the bath site have to be used as additional
variational parameters, but in the present case (half filling) their
value is fixed due to particle-hole symmetry.  Starting from high
temperature $T$ (low $\beta = T^{-1}$) and weak interaction $U$, one
can easily perform a global search to obtain a solution of
Eq.~(\ref{eq:DiffEulerEqn_eq}).  The full $T$--$U$ phase diagram is
then explored using a local search based on Broyden's
method,\citep{kelley1987,broyden1965} starting from the solution at a
nearby point in the phase diagram. By lowering the temperature $T$,
i.e., increasing $\beta$ , we find three solutions for certain values
of $U$: There is a metallic solution with large $V'_{\rm opt}$, which
is adiabatically connected to the metallic phase at weak $U$, an
insulating one with small $V'_{\rm opt}$, connected to the strong-$U$
limit, and a third solution with intermediate $V'_{\rm opt}$, which is
thermodynamically unstable (see below). This indicates a first-order
phase transition with coexisting metallic and insulating phases.

In Fig.~\ref{fig:eqVdocc}(b) we additionally present the double
occupancies for the respective optimal solutions. In the coexistence
region the double occupancy, like the optimal hybridization, has three
branches. The branch for which $\ev<\n{\ua}\n{\da}>$ increases with
increasing $U$ corresponds to the thermodynamical unstable solution,
because for this phase one has $\partial \ev<\n{\ua}\n{\da}>/\partial
U = \partial^2 \Omega(\beta,U)/\partial U^2 > 0$, which violates a
thermodynamical stability condition.  In fact, the system undergoes a
first-order phase transition at a critical interaction, the value of
which can be inferred from the Maxwell construction,\citep{strand2011}
as shown in the inset of Fig.~\ref{fig:eqVdocc}(b): The double
occupancy jumps at the critical interaction $U_{\rm c}$, determined by
requiring that the shaded areas on both sides of the jump be equal.
In addition, the lower and upper boundaries of the coexistence region,
$U_{{\rm c}1}$ and $U_{{\rm c}2}$, can be read off at the spinodal
points of the curve [see arrows in Fig.~\ref{fig:eqVdocc}(b)].

\begin{figure}[t]
  \centering
  \includegraphics[width=0.9\columnwidth]{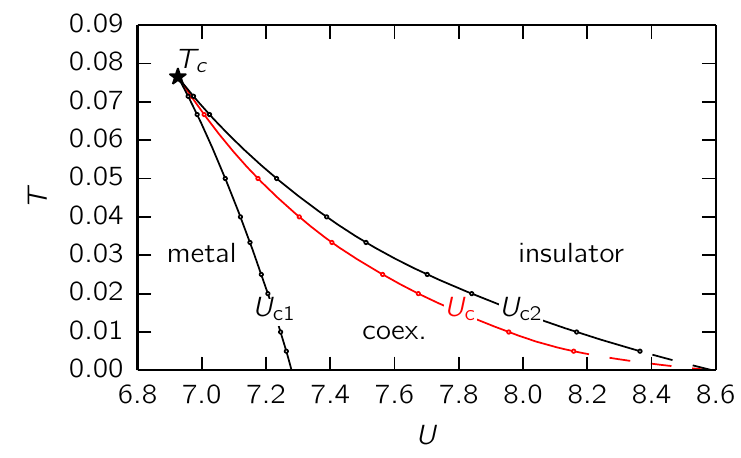}
  \caption{Phase diagram of the Mott transition in the half-filled
    Hubbard model on a one-dimensional lattice as obtained from the
    DIA with a two-site reference system. Below the critical
    temperature $T_c$ metallic solutions exist up to interactions
    $U\leq U_{{\rm c}2}$, insulating solutions exist down to $U\geq
    U_{{\rm c}1}$, and in between both coexist. Red line: first-order
    phase boundary $U_{\rm c}(T)$.}
  \label{fig:eqUc}
\end{figure}

Results for different temperatures are collected in the phase diagram
shown in Fig.~\ref{fig:eqUc}. Metallic and insulating solutions
coexist in a triangular-shaped region, bounded by the curves $U_{{\rm
    c}1}(T)$ and $U_{{\rm c}2}(T)$.  Within the coexistence region,
there is a line $U_{\rm c}(T)$ of first-order transitions terminating
in a second-order critical point at the temperature $T_{c}$.  For
temperatures above $T_{c}$ the Mott metal-insulator transition becomes
a smooth crossover. Extrapolating our data to $T=0$, we find $U_{{\rm
    c}1}(0) \approx 7.28 \approx 5.15\var$ and $U_{\rm c}(0) = U_{{\rm
    c}2}(0) \approx 8.59 \approx 6.07\var$, both of which fall within
a range of results obtained earlier for other lattices (where the
variance of the density of states has been used as the energy
unit).\citep{potthoff2003b} The value of $U_{{\rm c}2}(0)$ obtained
within the DIA for the Bethe lattice is in remarkably good agreement
with DMFT+NRG\citep{bulla1999} ($U_{\rm c}(0) = 5.88$).  The value
obtained for the critical temperature, $T_c\approx 0.077 \approx
0.054\var$, is more sensitive to the lattice geometry: For the
semi-elliptical DOS one finds $T_c\approx 0.03$ within the two-site
DIA.\citep{potthoff2003b} This value underestimates the DMFT result by
50\%, but quantitative agreement is obtained already by adding only
three bath sites.\citep{pozgajcic2004} 

As there is no Mott transition at a finite Hubbard-$U$ in the one-dimensional model, 
\citep{giamarchi2004} let us point out again that the DIA is a mean-field approach. 
It is therefore not really sensitive to the lattice dimension, and the results rather depend on the lattice geometry via the variance of the DOS only.
The one-dimensional case is studied here for purely technical reasons, and the results should be seen as representative for the model on higher-dimensional lattices.

We conclude that the present implementation of the two-site DIA
reproduces the known results for the Mott transition obtained earlier
where the phase diagram has been constructed from the explicit
calculation of the grand potential. The agreement between the results
of the two different numerical approaches is fully quantitative.
As compared to the full DMFT solution, the two-site
approximation qualitatively captures the correct topology of the
equilibrium phase diagram.  Quantitatively, $U_{{\rm c}2}$ is
predicted quite accurately while $U_{{\rm c}1}$ is over- and $T_{c}$
is underestimated.  For the present study, we will nevertheless
restrict ourselves to the two-site approximation since the
computational effort is considerably higher for the nonequilibrium
case.  More importantly, nonequilibrium calculations with more than a
single bath site are not easily stabilized numerically with the
present implementation. \citep{hofmann2015}

\section{Dynamical Mott transition}
\label{sec:dynamit}

In the following we will discuss the real-time dynamics of the Hubbard
model induced by sudden quenches or ramps of the interaction from a
``free'' initial state to arbitrary final interactions $U_{\rm
  fin}>0$.  Within the SFT the optimal parameters of the reference
system are undefined for a free system due to the vanishing
self-energy.  For practical reasons we therefore consider initial
states with $U_{\rm ini} = 0.01$.

We furthermore choose the initial inverse temperature $\beta = 10$.
Essentially, this corresponds to a zero-temperature initial state: As
concerns the reference system, there is hardly any change in the
optimal hybridization parameter with temperature in the limit $U\to
0$, as can be seen from Fig.~\ref{fig:eqVdocc}(a).  The remaining
temperature dependence via the noninteracting Green's function of the
original system [see Eq.\ (\ref{eq:SFTfcl})] is very weak for lower
temperatures.  Consequently, there is hardly any temperature
dependence seen in the nonequilibrium results.  This has been verified
numerically (up to $\beta\leq 40$).

The interaction is switched from $U_{\rm ini}$ to $U_{\rm fin}$ via
$U(t) = U_{\rm ini} + (U_{\rm fin} - U_{\rm ini})r(t)$ by either
quenching,
\begin{equation}
  r(t) = \Theta(t) \; , 
\end{equation}  
or conducting cosine-shaped ramps of different duration $\Delta t_{\rm
  ramp}$, i.e.,
\begin{equation}
  \label{eq:ramp}
  r(t) = (1-\cos(\pi t/\Delta
  t_{\rm ramp}))/2 \; . 
\end{equation}  
Both cases will be discussed successively in the next two subsections.

\subsection{Interaction quenches}
\label{sec:interaction-quenches}

\begin{figure}[t]
  \centering
  \includegraphics[width=0.9\columnwidth]{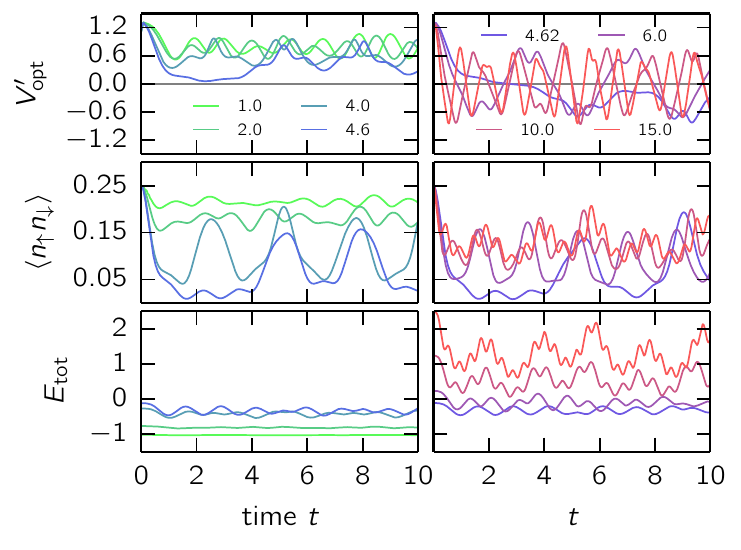}%
  \caption{Time dependencies of the optimal hybridization $V'_{\rm
      opt}$, the double occupancy $\ev<n_\ua n_\da>$, and the total
    energy $E_{\rm tot}$ for interaction quenches starting from
    $U_{\rm ini} = 0.01$ to different $U_{\rm fin}$ (see color
    labels). Left: weak-coupling regime, $U_{\rm fin} < U_{\rm c}^{\rm
      dyn}$. Right: strong-coupling regime, $U_{\rm fin} > U_{\rm
      c}^{\rm dyn}$.}
  \label{fig:weakstrong}
\end{figure}

\begin{figure}[t]
  \centering
  \includegraphics[width=0.9\columnwidth]{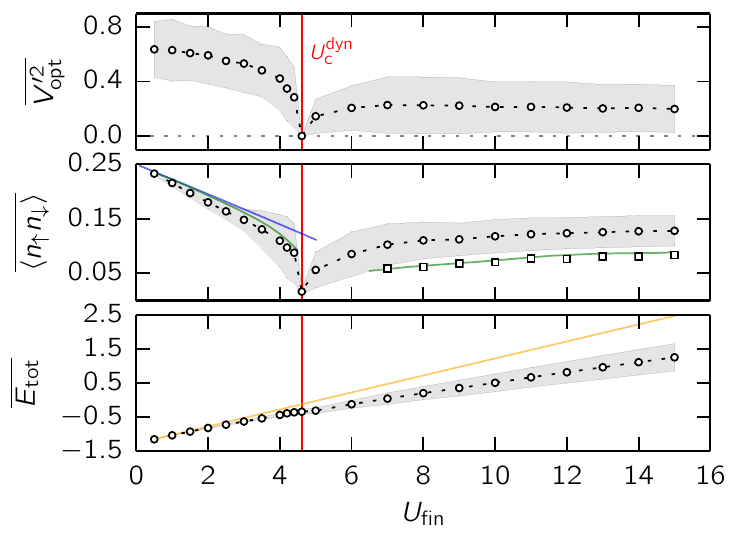}%
  \caption{Long-time averages (points) and fluctuations (shaded areas)
    of the optimal hybridization ${V'}^{2}_{\rm opt}$, the double
    occupancy $\ev<n_\ua n_\da>$ and the total energy $E_{\rm
      tot}$. Red lines: critical interaction $U_{\rm c}^{\rm dyn}$
    separating weak- and strong-coupling regime. Blue line:
    equilibrium values of the double occupancy at zero
    temperature. Green lines: thermal values of the double occupancy,
    which in the weak-coupling regime almost coincide with the
    long-time average, but in the strong-coupling regime match the
    minima of the double-occupancy oscillations (squares). Orange
    line: total energy right after the quench.  }
  \label{fig:av}
\end{figure}

Following the time evolution after a quench, we find two qualitatively
different response patterns for weak and strong final interactions,
which are well separated by a sharp transition point at a critical
interaction $U_{\rm c}^{\rm dyn}\approx 4.61$.  Results are presented
in Fig.~{\ref{fig:weakstrong}}, where we show the time dependence of
the optimal hybridization parameter, the double occupancy, and the
total energy. Moreover, in Fig.~{\ref{fig:av}} we show for each
time-dependent quantity $Q(t)$ the time average
\begin{equation}
  \overline Q = \lim_{t\ra\infty}\frac{1}{t}\int_0^t dt'\, Q(t')\,,
\end{equation}
and the fluctuations
\begin{equation}
  \overline {\Delta Q} = \overline{(Q-\overline Q)^2}^{\frac{1}{2}}\,.
\end{equation}

Let us first focus on \emph{weak} quenches, i.e., $U_{\rm fin} <
U_{\rm c}^{\rm dyn}$. For the optimal hybridization parameter $V'_{\rm
  opt}(t)$ we observe a quick drop to smaller values within
approximately one inverse hopping, followed by moderate oscillations
around some constant value, see Fig.~\ref{fig:weakstrong} (top left).
For final interactions $U_{\rm fin} \lesssim 4$, the long-time average
of the optimal hybridization slightly decreases with increasing
$U_{\rm fin}$ (Fig.~\ref{fig:av}, top). On the same short time scale,
the double occupancy decays from its noninteracting initial value,
i.e., the Coulomb repulsion quickly suppresses doubly occupied
sites. For final interactions $U_{\rm fin} \gtrsim 3$ we find a strong
initial drop and pronounced periodic recurrences.  However, these
recurrences shift to later and later times upon increasing $U_{\rm
  fin}$.  In addition, small regular oscillations around some value
close to zero become apparent, see Fig.~\ref{fig:weakstrong} (middle
left).

The exact value of the total energy right after the quench (at
$t=t_0^+$) is given by the expectation value of the Hamiltonian in the
noninteracting initial state, $E_{\rm tot}(t_0^+) = E_{\rm kin}(t_0) +
U_{\rm fin}/4$, which increases linearly with the final interaction.
For weak quenches we find that this value is relatively well
conserved, apart from a small drop of the time-averaged value (of less
than $0.1$), and some moderate oscillations of about $5\%$ or less for
increased quench size (up to $U_{\rm fin}\leq 4$) and when compared to
the respective long-time average (Fig.~\ref{fig:av}, bottom). By
comparison with a thermal ensemble for the same interaction $U_{\rm
  fin}$, i.e., by comparing with equilibrium two-site DIA
calculations, we can thus ascribe an effective temperature $T_{\rm
  eff}$ to the long-time averages by demanding that $\overline{E_{\rm
    tot}} = E_{\rm tot}^{\rm eq}(T_{\rm eff})$.  The effective
temperature increases from $T_{\rm eff} \approx 0.12$ for $U_{\rm fin}
= 1$ to $T_{\rm eff} \approx 0.28$ for $U_{\rm fin} = 4$.  The
corresponding thermal value for the double occupancy roughly agrees
with the respective time-averaged value (see Fig.~\ref{fig:av},
middle).  This is in agreement with the prethermalization scenario
\citep{moeckel2008,eckstein2009b} observed in DMFT calculations.

\begin{figure}[t]
  \centering
  \includegraphics[width=0.9\columnwidth]{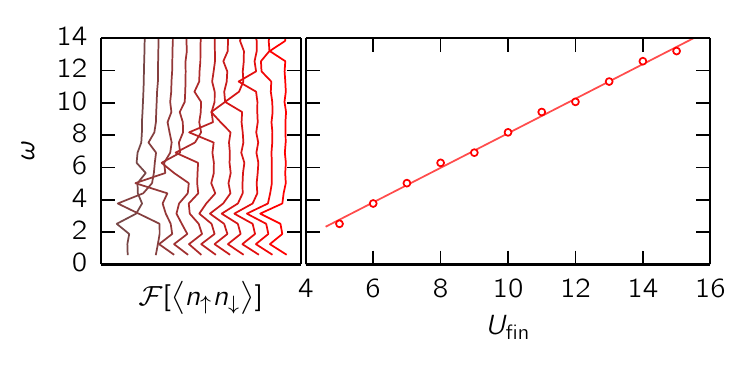}%
  \caption{Left: Fourier transform ($\mc F$) of the double occupancy.
    Plots have been shifted for better visibility. Right: linear
    dependence of the dominant frequency on the final interaction
    $U_{\rm fin}$ in the strong coupling regime $U_{\rm fin}>U_{\rm
      c}^{\rm dyn}$. Note the interaction-independent small beating
    frequencies (left).}
  \label{fig:quench_fft}
\end{figure}

We now turn to \emph{strong} quenches, i.e., $U_{\rm fin} > U_{\rm
  c}^{\rm dyn}$. Here the time-dependent behavior of the system
drastically differs from that in the regime of weak quenches. For
$U_{\rm fin} \gtrsim 6$ we find a quite regular oscillatory behavior
for all relevant quantities.  A Fourier analysis of the oscillations
in the double occupancy reveals that oscillations occur with
frequencies approximately given by $U_{\rm fin}$, as shown in
Fig.~\ref{fig:quench_fft}, i.e., the characteristic frequency for
collapse-and-revival oscillations in the atomic limit. On top of this,
there are slow beatings, which probably should be ascribed to
finite-size effects and which appear to be independent of the
interaction.  In the long-time limit, the double occupancy slowly
increases with $U_{\rm fin}$. However, it does not reach its free
value (i.e., $\ev<\n{\ua}\n{\da}>_0 = 0.25$) again, as perturbative
arguments would suggest,\citep{eckstein2009b} i.e., the two-site
approximation seems to underestimate the actual double occupancy in
the strong-coupling limit (see Fig.~\ref{fig:av}, middle).

The optimal hybridization parameter strongly oscillates around zero,
see Fig.~\ref{fig:weakstrong} (top right).  In equilibrium and for
strong interactions the quasiparticle weight (for a two-site system)
is given by $36V'^2/U^2$,\citep{lange1998} so that strong collapses
and revivals of the \emph{square} of the optimal parameter would
correspond to an oscillatory behavior of the Fermi-surface
discontinuity, as has been observed in DMFT
calculations. \citep{eckstein2009b} In Fig.~\ref{fig:av} (top) we
therefore show the long-time behavior of $V'^2_{\rm opt}(t)$.  With
increasing interactions $U_{\rm fin}$ we find that both its average
and its fluctuations quickly saturate.

For strong final interactions, conservation of total energy becomes
rather poor and, compared to the weak-coupling case, the time-averaged
value differs more significantly from the exact value. Nevertheless,
we may again compare the long-time average to an appropriate thermal
value to extract the effective temperature $T_{\rm eff}$.  As compared
to the weak-coupling regime, the effective temperatures are roughly an
order of magnitude higher and, apart from an offset, scale linearly
with $U_{\rm fin}$.  Interestingly, from this we find thermal values
of the double occupancy which in fact coincide with the overall minima
of its time-dependent oscillations, see Fig.~\ref{fig:av} (middle).
This is again in line with the DMFT results.\citep{eckstein2009b}

\begin{figure}[t]
  \centering
  \includegraphics[width=0.9\columnwidth]{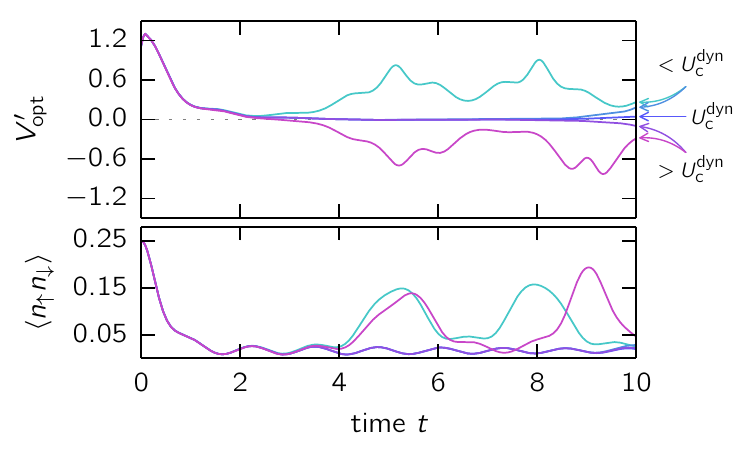}%
  \caption{Dynamical decoupling of the bath site at the critical point
    ($U_{\rm fin} = U_{\rm c}^{\rm dyn}$) (top) and the corresponding
    dynamics of the double occupancy (bottom). Additional curves:
    dynamics for final interactions differing by less than $0.3\%$
    from $U_{\rm c}^{\rm dyn}$.  Note the strong impact upon tiny
    changes of $U_{\rm fin}$ indicating a \emph{sharp} transition
    between the two regimes.}
  \label{fig:decouple}
\end{figure}

We now focus on the dynamics close to the \emph{critical} point
$U_{\rm c}^{\rm dyn}\approx 4.61$, see Fig.~\ref{fig:decouple}. In
this regime the behavior of $V_\text{opt}'(t)$ for $U_{\rm fin}
\lesssim U_{\rm c}^{\rm dyn}$ and $U_{\rm fin} \gtrsim U_{\rm c}^{\rm
  dyn}$ becomes very similar: Within two inverse hoppings, the optimal
hybridization strength decays almost to zero, but then revives to
positive values for $U_{\rm fin} \lesssim U_{\rm c}^{\rm dyn}$ and
shows slow oscillations with relatively large amplitude. The same
dynamics, but with opposite sign of the optimal parameter at long
times, is observed for $U_{\rm fin} \gtrsim U_{\rm c}^{\rm dyn}$.
This is accompanied by a decay of the double occupancy down to almost
zero ($\overline{\ev<n_{\ua}n_{\da}>} \approx 0.016$), followed by
strong revivals which are in phase with $V_\text{opt}'(t)$.  As
discussed for the weak-quench regime, these oscillations shift to
later and later times for quenches closer and closer to the critical
value.  Finally, right at the critical point, no revivals are
observed, i.e., the bath dynamically \emph{decouples}, and
$V_\text{opt}'(t)$ remains zero up to the longest simulated times.
For $U_{\rm fin} = U_{\rm c}^{\rm dyn}$ the double occupancy merely
shows weak oscillates around its long-time average.

We conclude that, within the two-site DIA, the dynamical Mott
transition is described as a sharp transition characterized by {\em
  critical} behavior in the $U_{\rm fin}$-dependence of the quantities
shown in Fig.~\ref{fig:av}.  One may speculate that in calculations
with more bath degrees of freedom in the DIA, some bath sites which
represent low energy degrees of freedom would decouple whereas others
would remain connected to the correlated impurity. Nevertheless, even
on the level of the two-site approximation, there is a surprisingly
good agreement of the critical interaction with results from the
DMFT\citep{eckstein2009b} ($U_{\rm c, DMFT}^{\rm dyn}\approx 3.2$)
and the Gutzwiller ansatz\citep{schiro2011} ($U_{\rm c, Gutzw}^{\rm
  dyn}\approx 3.3$) when comparing with the value rescaled by the
variance of the one-dimensional DOS, i.e., with $U_{\rm c}^{\rm
  dyn}\approx 4.61 \approx 3.26\var$.

Within the DMFT \citep{eckstein2009b} a rapid thermalization is found
at $U_{\rm fin} = U_{\rm c}^{\rm dyn}$, and the thermalized state is
characterized as a bad metal.  Opposed to this, within the two-site
DIA, a complete decoupling of the bath site right at the critical
point implies that the final state is described on the level of the
Hubbard-I approximation. \citep{hubbard1963} Note that therewith the
dynamical Mott transition is very similar to the equilibrium Mott
transition at zero temperature which is also characterized by a
vanishing hybridization to the bath site.  In both cases, the
Hubbard-I approximation must be seen as a comparatively crude
description of the bad metal or Mott insulator, respectively, and one
cannot expect a fully consistent picture on this level.  In the
nonequilibrium case, for example, the determination of the effective
temperature by comparison with equilibrium two-site DIA calculations
via $\overline{E_{\rm tot}} = E_{\rm tot}^{\rm eq}(T_{\rm eff})$
yields $T_{\rm eff} \approx 0.3$.  The resulting thermal double
occupancy of $\ev<\n{\ua}\n{\da}> \approx 0.1$, however, turns out too
large as compared with the time average
$\overline{\ev<n_{\ua}n_{\da}>} \approx 0.016$.  A better agreement is
found when estimating $T_{\rm eff}$ by comparing with the Hubbard-I
solution, where the bath site is decoupled.  This yields $T_{\rm eff}
\approx 0.6$ and $\ev<\n{\ua}\n{\da}> \approx 0.02$.  However, at
finite temperatures, the Hubbard-I solution is only metastable.  One
may summarize that for the final state at $U_{\rm fin} = U_{\rm
  c}^{\rm dyn}$, our findings more resemble the predictions of the
Gutzwiller approach \citep{schiro2011} rather than those of the DMFT.

It is also instructive to interpret our results in the context of a
recent proposal \citep{schiro2011,zitko2015} stating that the metallic
phase of the paramagnetic and particle-hole symmetric Hubbard model in
infinite dimensions can be seen as a phase with a spontaneously broken
local ${\mathbb Z}_{\mathrm{2}}$ gauge symmetry. At the
zero-temperature transition to the Mott insulating phase, the
${\mathbb Z}_{\mathrm{2}}$ symmetry is restored. This picture of the
Mott transition is nicely captured by the mean-field theory of a
slave-spin model, \citep{medici2005b,rueegg2010} which is essentially
equivalent to the Gutzwiller approach. Using the numerical
renormalization group, it has been demonstrated \citep{zitko2015} that
the ${\mathbb Z}_{\mathrm{2}}$ symmetry breaking at the $T=0$ Mott
critical point takes place in the full model and is not an artifact of
the mean-field approach. The slave-spin magnetization can serve as an
order parameter for the metal-insulator transition.

Within the DIA, the optimal hybridization $V_\text{opt}'$ with the
auxiliary bath site at zero one-particle energy can be seen as an
analogous order parameter.  There is, however, an important difference
since the corresponding symmetry group is U(1): Physical observables
depend on $|V_\text{opt}'|^{2}$ only and are therefore invariant under
a phase change of $V_\text{opt}'$.  The fact that $V_\text{opt}' \ne
0$ in the metallic phase can thus be interpreted as a spontaneous
breaking of a local U(1) gauge symmetry which is restored, i.e.,
$V_\text{opt}' = 0$, in the Mott insulator at $T=0$.  The choice
$V_\text{opt}' > 0$ for the metallic phase just fixes the gauge.

Our numerical results for the quench dynamics at $U_{\rm fin} = U_{\rm
  c}^{\rm dyn}$ indicate that there is a transition from a symmetry
broken, $V_\text{opt}'(t=0) \ne 0$ state at time $t=0$ to a symmetric
state, $V_\text{opt}'(t) = 0$ for $t \to \infty$, i.e., a
time-dependent Mott transition -- similar to that in the Gutzwiller
approach. \citep{schiro2011} In the Gutzwiller approach and in the
two-site DIA as well, the final state that is reached for $t \to
\infty$ is not the thermal state.  Namely, a full two-site DIA
equilibrium calculation for $U=U_{\rm c}^{\rm dyn}$ would give
$V_\text{opt}' \ne 0$ at any finite temperature, and the corresponding
thermal state has a lower grand potential than that of the
Hubbard-I-like state which is obtained in the equilibrium calculation
by {\em ad hoc} setting $V_\text{opt}'=0$ (which is always a
stationary point of the grand potential).  Hence, within the two-site
DIA restoring the local U(1) gauge symmetry in the time-dependent Mott
transition necessarily implies that the final state is nonthermal.
Within the DMFT, in contrast, the final state, which is (rapidly)
reached after a quench to $U_{\rm fin} = U_{\rm c}^{\rm dyn}$, is a
thermal state with a high temperature (more than an order of magnitude
higher than $T_c$), \citep{eckstein2009b} which does not break the
${\mathbb Z}_{\mathrm{2}}$ symmetry. \citep{zitko2015} One may
speculate that within the DIA the hybridization of the zero-energy
mode with the impurity also vanishes in the thermal state at $T>0$ if
more and more bath sites are added, so that a rapid decoupling of this
mode at $U_{\rm fin} = U_{\rm c}^{\rm dyn}$ would not be at odds with
rapid thermalization.

\begin{figure}[t]
  \centering
  \includegraphics[width=0.9\columnwidth]{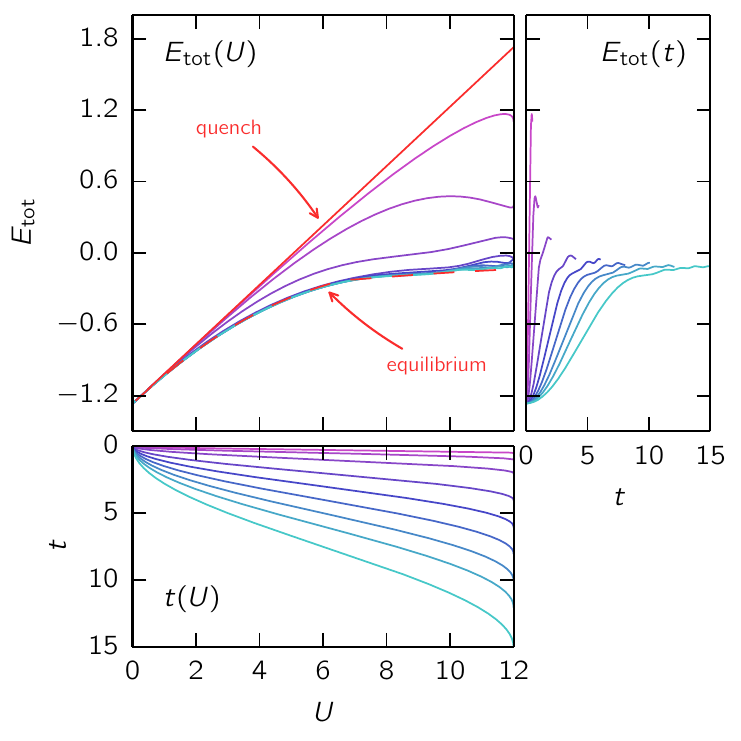}%
  \caption{Check of adiabaticity for ramps of the interaction from
    $U_{\rm ini} = 0.01$ to $U_{\rm fin} = 12$ with different ramp
    times ($\Delta t_{\rm ramp}\in \{0.5,1,2,4,6,8,10,12,15\}$,
    colored from purple to cyan). The total energy during the ramp is
    plotted as a function of the instantaneous interaction (top left),
    i.e., $E_{\rm tot}(U) \equiv E_{\rm tot}(t(U))$, as obtained from
    the inverse of (cosine) ramp profile $U(t)$, Eq.~(\ref{eq:ramp}),
    (bottom) and from the time-dependent total energy $E_{\rm tot}(t)$
    (right). The results are contrasted with the total energy after a
    quench (straight red line) and the equilibrium energy at the same
    final interaction (dashed red line).}
  \label{fig:EtU_adiabatic}
\end{figure}

\begin{figure}[t]
  \centering
  \includegraphics[width=0.9\columnwidth]{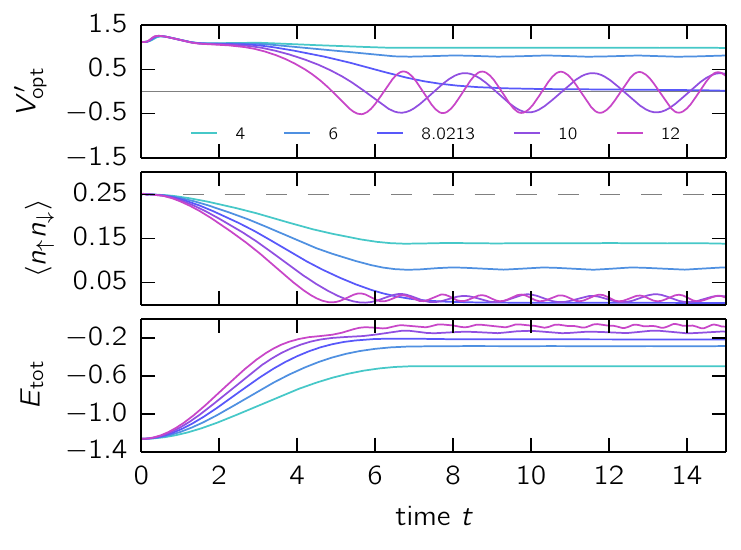}%
  \caption{Time dependencies of the optimal hybridization $V'_{\rm
      opt}$, the double occupancy $\ev<n_\ua n_\da>$ and the total
    energy $E_{\rm tot}$ (from top to bottom) for ramps of the
    interaction with $\Delta t_{\rm ramp} = 7$, starting from $U_{\rm
      ini} = 0.01$ to different $U_{\rm fin}$ (as indicated in the top
    panel) in the weak- and the strong-coupling regime as well as
    right at the dynamical critical point.}
  \label{fig:ramp7}
\end{figure}

\subsection{Ramps of the interaction}
\label{sec:ramps-interaction}

\begin{figure*}[t]
  \centering \subfloat[$U_{\rm fin} < U_{\rm c}^{\rm
    dyn}$]{\label{fig:nkt_ramp7_below}\includegraphics[width=.5\columnwidth]{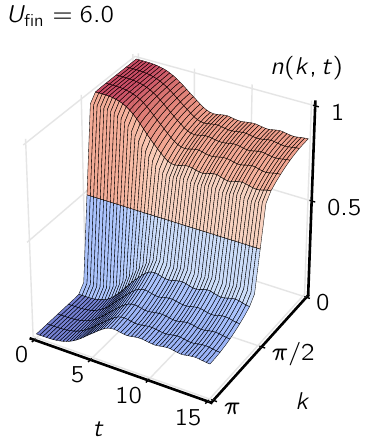}}%
  \subfloat[$U_{\rm fin} > U_{\rm c}^{\rm
    dyn}$]{\label{fig:nkt_ramp7_above}\includegraphics[width=.5\columnwidth]{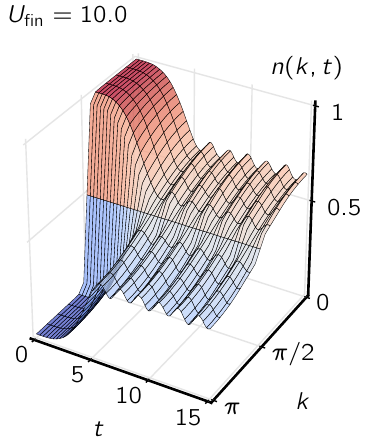}}%
  \subfloat[$U_{\rm fin} = U_{\rm c}^{\rm
    dyn}$]{\label{fig:nkt_ramp7_crit}\includegraphics[width=.5\columnwidth]{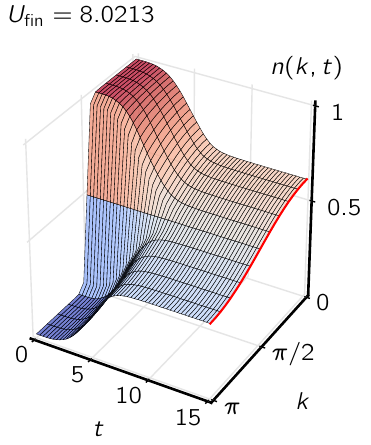}}%
  \subfloat[relative error for
  fits]{\label{fig:nkt_ramp7_fit}\includegraphics[width=.5\columnwidth]{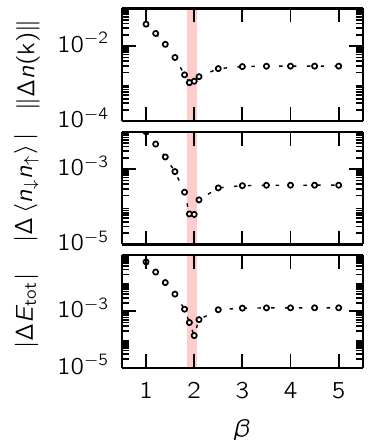}}%
  \caption{Time dependent momentum distribution for a ramp with
    $\Delta t_{\rm ramp} = 7$ ending at different interactions
    {\protect\subref{fig:nkt_ramp7_below}} below,
    {\protect\subref{fig:nkt_ramp7_above}} above, and
    {\protect\subref{fig:nkt_ramp7_crit}} right at the critical
    point. Precise numbers are given in the top left corner of each
    plot. For the latter the long-time average of the momentum
    distribution is fitted with an equilibrium distribution within the
    Hubbard-I approximation (red line). Relative errors of fits for
    the momentum distribution, the double occupancy, and the total
    energy (from top to bottom) at different temperatures are shown in
    {\protect\subref{fig:nkt_ramp7_fit}}; an error estimate for the
    effective temperature $T_{\rm eff} \approx 0.5$ is indicated by a
    red shaded area.}
  \label{fig:nkt_ramp7}
\end{figure*}

The previous discussion provokes the question whether the dynamical
Mott transition and the conventional equilibrium Mott transition can
be smoothly connected to each other.  Since the dynamical transition
occurs at a much weaker interaction, it is not at all obvious whether
the two phenomena are related at all.  One route to study this
question is to consider a ramp with a finite duration rather than an
instantaneous quench of the interaction as has been done using the
Gutzwiller approach in Ref.\ \onlinecite{sandri2012}.  Ramping the
interaction in a short time from $U_{\rm ini}$ to $U_{\rm fin}$ will
make contact to the results found for a sudden quench.  On the other
hand, for ramps with infinite duration, i.e., if the interaction is
changed adiabatically rather than suddenly, the system evolves along
paths within the equilibrium phase diagram and will cross the line of
equilibrium transitions (see Fig.~\ref{fig:eqUc}).  In fact, assuming
that there is a critical interaction for any ramp time $\Delta t_{\rm
  ramp}$ at all, one should expect that, with increasing $\Delta
t_{\rm ramp}$, the critical interaction crosses over from $U_{\rm
  c}^{\rm dyn} \approx 4.61$ (sudden quench) to $U_{{\rm c}2} \approx
8.59$ ($T=0$), since starting from a zero-temperature initial state,
an adiabatic process will result in a zero-temperature final state.

To test our expectation we therefore consider a sequence of
(cosine-shaped) ramps with different duration $\Delta t_{\rm ramp}$,
see Eq.~(\ref{eq:ramp}).  Here we are limited to finite propagation
times, $t_{\rm max} \leq 25$, for practical reasons.  Nevertheless,
this allows us to study the relevant critical behavior for ramp times
up to $\Delta t_{\rm ramp} \leq 20$.

We begin the discussion for ramps of different duration to the same
final interaction $U_{\rm fin} = 12$, starting from the same initial
state that has also been considered for the quenches discussed in
Sec.~\ref{sec:interaction-quenches}.  As a measure of adiabaticity
of the process we compare the total energy as a function of the
interaction \emph{during} the ramp with the corresponding equilibrium
result.  This is shown in Fig.~\ref{fig:EtU_adiabatic} where the
time-dependent value of the total energy $E_{\rm tot}(t)$ during the
ramp is plotted against the instantaneous value $U(t)$ of the
interaction.  The resulting function $E_{\rm tot}(U)$ can be compared
with the equilibrium total energy (dashed line) as well as with the
total energy after a sudden quench (straight line).  With increasing
ramp duration the curves $E_{\rm tot}(U)$ converge to the equilibrium
result, i.e., to the $U$-dependence of the ground-state energy.  For
ramp times $\Delta t_{\rm ramp} \gtrsim 15$, the process is almost
perfectly adiabatic.

As is discussed in the context of the (quantum) Kibble-Zurek
mechanism, \citep{dziarmaga2010,francuz2016} the dependence of the
energy difference $\Delta E \equiv E_{\rm tot}(\Delta t_{\rm ramp}) -
E_{\rm tot}(\infty)$ on the ramp time $\Delta t_{\rm ramp}$ should
asymptotically follow an inverse power law.
Figure~\ref{fig:EtU_adiabatic} demonstrates that the excitation energy
does decrease with increasing $\Delta t_{\rm ramp}$. Here, one would
expect the mean-field exponent, i.e., $\Delta E \propto 1/\Delta
t_{\rm ramp}$ (see Ref.~\onlinecite{sandri2012} for a discussion),
and, roughly, our data are in fact consistent with this
expectation. To extract a reliable value for the exponent, however,
calculations for much longer ramp times would be necessary.

Independent of the ramp duration, we find essentially the same
distinction between a weak- and a strong-coupling regime that has been
discussed for the case of a sudden quench.  The two regimes are
sharply separated by a critical interaction $U_{\rm c}^{\rm dyn} =
U_{\rm c}^{\rm dyn}(\Delta t_{\rm ramp})$ which depends on the ramp
duration (see discussion below). With increasing ramp time the
dynamics becomes more well behaved in the sense that energy
conservation becomes almost perfect for weak final interactions and is
strongly improved in the strong-coupling case.  As an example, in
Fig.~\ref{fig:ramp7}, we show the time dependencies of the optimal
hybridization, of the double occupancy and of the total energy for
different values of $U_{\rm fin}$ which are located below, right at,
and above the dynamical critical interaction $U_{\rm c}^{\rm dyn}
\approx 8.02$ for a ramp with $\Delta t_{\rm ramp} = 7$.  Note that
the process is clearly nonadiabatic.  For $U_{\rm fin} \leq U_{\rm
  c}^{\rm dyn}$ we find that there are almost no oscillations of the
optimal hybridization parameter after the ramp is completed.  The same
holds true for the double occupancy.  Its (time-averaged) value after
the ramp only slightly increases with increasing final interaction
after having reached a minimum (close to zero) right at the critical
point.  This already indicates proximity to an adiabatic process where
the double occupancy would just follow its equilibrium value, i.e.,
where it would monotonically decrease with increasing final
interactions [see Fig.~\ref{fig:eqVdocc}(b)].

We also obtain reasonable results for the time-dependent momentum
distribution $n(k,t)$ and, contrary to the study of quench dynamics,
can therefore more comprehensively focus on the question of
thermalization.  In Fig.~\ref{fig:nkt_ramp7}, we show three different
examples for the final-state dynamics of $n(k,t)$, exemplary for the
weak- [Fig.~\ref{fig:nkt_ramp7_below}] and for the strong-coupling
case [Fig.~\ref{fig:nkt_ramp7_above}] as well as for $U_{\rm fin} =
U_{\rm c}^{\rm dyn}$ [Fig.~\ref{fig:nkt_ramp7_crit}].  We again
consider ramps with $\Delta t_{\rm ramp} = 7$.  The initial state is
characterized by a sharp Fermi-surface discontinuity, which is
slightly washed out by the finite temperature ($\beta = 10$).  The
final state that is reached in the long-time limit either shows a
sharp jump of $n(k,t)$ at the Fermi surface (in case of $U_{\rm fin} <
U_{\rm c}^{\rm dyn}$) or collapse-and-revival oscillations ($U_{\rm
  fin} > U_{\rm c}^{\rm dyn}$).  This is very similar to the DMFT
results in the quench case.\citep{eckstein2009b} Right at the critical
point ($U_{\rm fin} = U_{\rm c}^{\rm dyn}$) we also find fast
thermalization toward a hot thermal distribution immediately after the
ramp is completed.  Comparing with Hubbard-I equilibrium calculations,
we find an effective temperature of $T_{\rm eff} \approx 0.5$, see
Figs.~\ref{fig:nkt_ramp7_crit} and \ref{fig:nkt_ramp7_fit}.  Note that
this is somewhat lower than the effective temperature that had been
obtained for the quench ($T_{\rm eff} \approx 0.6$).

\begin{figure}[t]
  \centering
  \includegraphics[width=0.9\columnwidth]{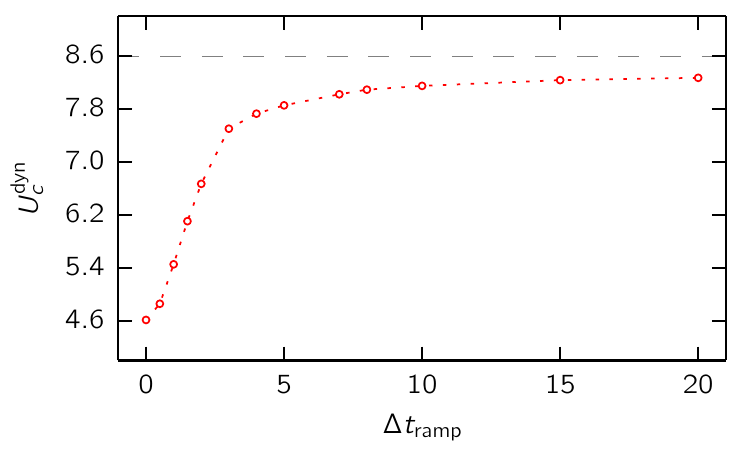}%
  \caption{Dynamical critical interaction $U_{\rm c}^{\rm dyn}$ as a
    function of the ramp time $\Delta t_{\rm ramp}$ determined for
    cosine-shaped ramps starting from $U_{\rm ini} = 0.01$. The
    equilibrium value $U_{{\rm c}2} \approx 8.59$ for the critical
    interaction at zero temperature is indicated by a gray dashed
    line.}
  \label{fig:Ucdyns_dtramps}
\end{figure}

Let us finally come back to the original motivation to study ramps of
the interaction.  We consider ramps with various durations bridging
the limit of an instantaneous quench $\Delta t_{\rm ramp}=0$ and the
adiabatic limit $\Delta t_{\rm ramp}\to \infty$. For each ramp time,
we have performed a series of calculations with different $U_{\rm
  fin}$ to extract the respective value of the dynamical critical
interaction $U_{\rm c}^{\rm dyn}$.  The latter is indeed well defined
in the whole $\Delta t_{\rm ramp}$ regime.  Its dependence on the ramp
time for $\Delta t_{\rm ramp} \leq 20$ is shown in
Fig.~\ref{fig:Ucdyns_dtramps}.

$U_{\rm c}^{\rm dyn}$ monotonically increases with $\Delta t_{\rm
  ramp}$ and seems to approach the value of the critical interaction
$U_{{\rm c}2} \approx 8.59$ for the zero-temperature Mott transition,
as obtained by the two-site DIA (cf. Sec.~\ref{sec:equil-mott-trans}).
However, the convergence turns out to be very slow.  We also cannot
fully exclude that the low but nonzero initial-state temperature has
some effect on the result expected for $\Delta t_{\rm ramp}\to \infty$
and that, even for a perfectly adiabatic process, the final-state
effective temperature becomes nonzero which would imply that $U_{\rm
  c}^{\rm dyn}$ converges to a somewhat lower value.  Nevertheless,
the results indeed clearly indicate that the dynamical Mott transition
and the equilibrium Mott transition are related phenomena which are
smoothly connected -- at least within the two-site DIA.  The same
conclusion can be drawn from the results of the Gutzwiller
calculations \citep{sandri2012} which, however, show additional
oscillations of the critical interaction when increasing the ramp
duration.  This effect is absent in the two-site DIA.

\subsection{Discussion of the method}
\label{sec:discussion}

To conclude this section, let us contrast our approach with
Hamiltonian-based methods which strive to solve the effective impurity
model of nonequilibrium DMFT exactly by mapping it onto a
single-impurity Anderson model with a {\em finite} number of bath
sites,\citep{gramsch2013} which can then be treated numerically. The
number of bath orbitals is systematically increased until the
properties of the DMFT bath are accurately represented. In the current
implementation, the number of bath sites needed for an accurate
representation of the DMFT bath scales roughly linearly with the
maximum simulation time (and it also weakly depends on the parameter
regime). This limits simulations to short times.  Different
implementations have been put forward to solve the finite impurity
model, using exact-diagonalization techniques, \citep{gramsch2013} the
multiconfiguration time-dependent Hartree method, \citep{balzer2015} as
well as an approach based on the matrix-product state
representation.\citep{wolf2014} With exact-diagonalization
methods\citep{gramsch2013} propagation up to $t_{\rm max} \approx 3$
inverse hoppings has been possible by providing $L_b = 8$ bath sites
at weak interactions, whereas using matrix-product states,
\citep{wolf2014} $t_{\rm max} \approx 7$ ($t_{\rm max} \approx 5.5$)
could be reached with $L_b = 24$ ($L_b = 18$) sites at strong (weak)
interactions, i.e., the Hamiltonian based solvers are currently aiming
at a numerical exact solution at short times.

In contrast to this, the self-energy functional approach maps the
original lattice-fermion problem onto an auxiliary model with a fixed,
small number of bath sites and, in the case of the dynamical impurity
approximation, a single correlated site.  Rather than aiming at an
exact solution of the nonequilibrium DMFT equations, the SFT provides
an independent variational scheme to determine the time-dependent
one-particle parameters of the reference system which only in the
limit of an infinite number of bath sites recovers the DMFT.
Formally, a qualitatively correct time evolution on much longer
propagation times is thus possible with a very small number of bath
sites. The present study has in fact shown that even with the most
simple reference system (with a single bath site only) one can make
close contact with full DMFT results.  The agreement between the
two-site DIA and full DMFT is qualitatively satisfying and close to
the critical point for the (dynamical) Mott transition even
quantitative.  This demonstrates that much of the essential physics
can be captured with a single time-dependent variational parameter.

The general framework of the SFT ensures that variational
approximations are conserving with respect to the particle number and
spin.  The possible violation of energy conservation, however, must be
seen as a major drawback of the present implementation of the DIA.
Ways to overcome this problem have been discussed in Ref.\
\onlinecite{hofmann2013}.  Here, we could show that energy
conservation is in fact violated but that, on the other hand, this
violation is moderate in the weak-coupling limit after a quantum
quench and even for strong interactions does not generally invalidate
the results which still agree qualitatively with DMFT.  Furthermore,
if the dynamics is initiated by ramping the interaction, energy
conservation is respected to a much higher degree.

\section{Summary}
\label{sec:conclusion}

The nonequilibrium extension of the self-energy functional theory
(SFT) has been applied to study the real-time dynamics of the
Fermi-Hubbard model initiated by sudden quenches and by ramps of the
interaction parameter, starting from the noninteracting limit to
different final values $U_{\rm fin}$.  As the simplest nontrivial
approximation within the SFT, which provides a local trial
self-energy, we have employed the two-site dynamical impurity
approximation (DIA).  The dynamical Mott metal-insulator transition
represents an ideal first test case for this method.

We have studied the dynamical Mott transition by systematically
tracing the time evolutions of the double occupancy, the total energy,
the momentum distribution, and the optimal hybridization parameter of
the reference system for different $U_{\rm fin}$.  All quantities are
found to exhibit distinct response behavior in the weak- and in the
strong-coupling regime.  Within the two-site DIA these regimes are
separated by a sharp critical interaction $U_{\rm c}^{\rm dyn}$ at
which the low-energy bath site decouples from the correlated site in
the course of time.  By analyzing long-time averages and comparing
these with thermal results, we have found fast thermalization for
quenches to $U_{\rm fin} = U_{\rm c}^{\rm dyn}$ and clear indications
for prethermalization in both, the weak- and the strong-coupling
regime.  In all relevant aspects, this is in surprisingly good
qualitative agreement with a previous nonequilibrium DMFT
study. \citep{eckstein2009b} This also holds for the numerical value
of the dynamical critical interaction which turns out to be roughly a
factor of two smaller than the critical value $U_{{\rm c}2}(0)$ for
the equilibrium Mott transition at zero temperature.

Comparing results for ramps of different duration, we could trace the
critical behavior in the whole range from a sudden quench to the limit
of an adiabatic quasistatic process.  We found that there is a
well-defined critical interaction $U_{\rm c}^{\rm dyn}$ in all cases
which monotonically increases with the ramp time and which converges
to the zero-temperature critical point $U_{{\rm c}2}$ in the adiabatic
limit.  Qualitatively, this agrees well with the predictions of the
Gutzwiller approach. \citep{schiro2011,sandri2012}

In view of the comparatively simple but successful two-site dynamical
impurity approximation it appears very promising to improve the
approach by considering more complex reference systems. An improved
study of the mean-field dynamics using a reference system with more
bath degrees of freedom suggests itself.  Furthermore, cluster
approximations generating non-local trial self-energies appear highly
interesting. Both routes are computationally demanding
\citep{hofmann2015} concerning both complexity but also stability, but
on the other hand also very promising.  Furthermore, the
nonequilibrium self-energy functional theory is a completely general
approach and can be applied to more complex systems beyond the
single-band Hubbard model at half filling -- similar to the
equilibrium case. \citep{potthoff2012,potthoff2014} We expect that the
method can serve as a highly useful tool to uncover and understand
intriguing phenomena of strongly-correlated many-body systems out of
equilibrium.

\acknowledgments

We would like to thank Christian Gramsch for numerous helpful
discussions. Support of this work by the Deutsche
Forschungsgemeinschaft (Germany) within the Sonderforschungsbereich
925 (projects B5 and B4) is gratefully acknowledged.


\def\BibitemShut#1{}


\begin{thebibliography}{55}%
  \makeatletter \providecommand \@ifxundefined [1]{%
    \@ifx{#1\undefined} }%
  \providecommand \@ifnum [1]{%
    \ifnum #1\expandafter \@firstoftwo \else \expandafter
    \@secondoftwo \fi }%
  \providecommand \@ifx [1]{%
    \ifx #1\expandafter \@firstoftwo \else \expandafter \@secondoftwo
    \fi }%
  \providecommand \natexlab [1]{#1}%
  \providecommand \enquote [1]{``#1''}%
  \providecommand \bibnamefont [1]{#1}%
  \providecommand \bibfnamefont [1]{#1}%
  \providecommand \citenamefont [1]{#1}%
  \providecommand \href@noop [0]{\@secondoftwo}%
  \providecommand \href [0]{\begingroup \@sanitize@url \@href}%
  \providecommand \@href[1]{\@@startlink{#1}\@@href}%
  \providecommand \@@href[1]{\endgroup#1\@@endlink}%
  \providecommand \@sanitize@url [0]{\catcode `\\12\catcode
    `\$12\catcode `\&12\catcode `\#12\catcode `\^12\catcode
    `\_12\catcode `\%12\relax}%
  \providecommand \@@startlink[1]{}%
  \providecommand \@@endlink[0]{}%
  \providecommand \url [0]{\begingroup\@sanitize@url \@url }%
  \providecommand \@url [1]{\endgroup\@href {#1}{\urlprefix }}%
  \providecommand \urlprefix [0]{URL }%
  \providecommand \Eprint [0]{\href }%
  \providecommand \doibase [0]{http://dx.doi.org/}%
  \providecommand \selectlanguage [0]{\@gobble}%
  \providecommand \bibinfo [0]{\@secondoftwo}%
  \providecommand \bibfield [0]{\@secondoftwo}%
  \providecommand \translation [1]{[#1]}%
  \providecommand \BibitemOpen [0]{}%
  \providecommand \bibitemStop [0]{}%
  \providecommand \bibitemNoStop [0]{.\EOS\space}%
  \providecommand \EOS [0]{\spacefactor3000\relax}%
  \providecommand \BibitemShut [1]{\csname bibitem#1\endcsname}%
  \let\auto@bib@innerbib\@empty
\bibitem [{\citenamefont {Imada}\ \emph {et~al.}(1998)\citenamefont
    {Imada}, \citenamefont {Fujimori},\ and\ \citenamefont
    {Tokura}}]{imada1998}%
  \BibitemOpen \bibfield {author} {\bibinfo {author} {\bibfnamefont
      {M.}~\bibnamefont {Imada}}, \bibinfo {author} {\bibfnamefont
      {A.}~\bibnamefont {Fujimori}}, \ and\ \bibinfo {author}
    {\bibfnamefont {Y.}~\bibnamefont {Tokura}},\ }\href {\doibase
    10.1103/RevModPhys.70.1039} {\bibfield {journal} {\bibinfo
      {journal} {Rev. Mod. Phys.}\ }\textbf {\bibinfo {volume} {70}},\
    \bibinfo {pages} {1039} (\bibinfo {year} {1998})}\BibitemShut
  {NoStop}%
\bibitem [{\citenamefont {J\"ordens}\ \emph
    {et~al.}(2008)\citenamefont {J\"ordens}, \citenamefont
    {Strohmaier}, \citenamefont {G\"unter}, \citenamefont {Moritz},\
    and\ \citenamefont {Esslinger}}]{joerdens2008}%
  \BibitemOpen \bibfield {author} {\bibinfo {author} {\bibfnamefont
      {R.}~\bibnamefont {J\"ordens}}, \bibinfo {author} {\bibfnamefont
      {N.}~\bibnamefont {Strohmaier}}, \bibinfo {author}
    {\bibfnamefont {K.}~\bibnamefont {G\"unter}}, \bibinfo {author}
    {\bibfnamefont {H.}~\bibnamefont {Moritz}}, \ and\ \bibinfo
    {author} {\bibfnamefont {T.}~\bibnamefont {Esslinger}},\ }\href
  {\doibase 10.1038/nature07244} {\bibfield {journal} {\bibinfo
      {journal} {Nature}\ }\textbf {\bibinfo {volume} {455}},\
    \bibinfo {pages} {204} (\bibinfo {year}
    {2008})}
  \BibitemShut {NoStop}%
\bibitem [{\citenamefont {Schneider}\ \emph
    {et~al.}(2008)\citenamefont {Schneider}, \citenamefont
    {Hackerm\"{u}ller}, \citenamefont {Will}, \citenamefont {Best},
    \citenamefont {Bloch}, \citenamefont {Costi}, \citenamefont
    {Helmes}, \citenamefont {Rasch},\ and\ \citenamefont
    {Rosch}}]{schneider2008}%
  \BibitemOpen \bibfield {author} {\bibinfo {author} {\bibfnamefont
      {U.}~\bibnamefont {Schneider}}, \bibinfo {author} {\bibfnamefont
      {L.}~\bibnamefont {Hackerm\"{u}ller}}, \bibinfo {author}
    {\bibfnamefont {S.}~\bibnamefont {Will}}, \bibinfo {author}
    {\bibfnamefont {T.}~\bibnamefont {Best}}, \bibinfo {author}
    {\bibfnamefont {I.}~\bibnamefont {Bloch}}, \bibinfo {author}
    {\bibfnamefont {T.~A.}\ \bibnamefont {Costi}}, \bibinfo {author}
    {\bibfnamefont {R.~W.}\ \bibnamefont {Helmes}}, \bibinfo {author}
    {\bibfnamefont {D.}~\bibnamefont {Rasch}}, \ and\ \bibinfo
    {author} {\bibfnamefont {A.}~\bibnamefont {Rosch}},\ }\href
  {\doibase 10.1126/science.1165449} {\bibfield {journal} {\bibinfo
      {journal} {Science}\ }\textbf {\bibinfo {volume} {322}},\
    \bibinfo {pages} {1520} (\bibinfo {year} {2008})}\BibitemShut
  {NoStop}%
\bibitem [{\citenamefont {Giamarchi}(2004)}]{giamarchi2004}%
  \BibitemOpen \bibfield {author} {\bibinfo {author} {\bibfnamefont
      {T.}~\bibnamefont {Giamarchi}},\ }\href@noop {} {\emph {\bibinfo
      {title} {Quantum Physics in One Dimension}}}\ (\bibinfo
  {publisher} {Oxford University Press},\ \bibinfo {year}
  {2004})\BibitemShut {NoStop}%
\bibitem [{\citenamefont {Bloch}\ \emph {et~al.}(2008)\citenamefont
    {Bloch}, \citenamefont {Dalibard},\ and\ \citenamefont
    {Zwerger}}]{bloch2008}%
  \BibitemOpen \bibfield {author} {\bibinfo {author} {\bibfnamefont
      {I.}~\bibnamefont {Bloch}}, \bibinfo {author} {\bibfnamefont
      {J.}~\bibnamefont {Dalibard}}, \ and\ \bibinfo {author}
    {\bibfnamefont {W.}~\bibnamefont {Zwerger}},\ }\href {\doibase
    10.1103/RevModPhys.80.885} {\bibfield {journal} {\bibinfo
      {journal} {Rev. Mod. Phys.}\ }\textbf {\bibinfo {volume} {80}},\
    \bibinfo {pages} {885} (\bibinfo {year} {2008})}
  \BibitemShut {NoStop}%
\bibitem [{\citenamefont {Lewenstein}\ \emph
    {et~al.}(2012)\citenamefont {Lewenstein}, \citenamefont
    {Sanpera},\ and\ \citenamefont {Ahufinger}}]{lewenstein2012}%
  \BibitemOpen \bibfield {author} {\bibinfo {author} {\bibfnamefont
      {M.}~\bibnamefont {Lewenstein}}, \bibinfo {author}
    {\bibfnamefont {A.}~\bibnamefont {Sanpera}}, \ and\ \bibinfo
    {author} {\bibfnamefont {V.}~\bibnamefont {Ahufinger}},\ }\href
  {\doibase 10.1093/acprof:oso/9780199573127.001.0001} {\emph
    {\bibinfo {title} {Ultracold Atoms in Optical Lattices: Simulating
        quantum many-body systems}}}\ (\bibinfo {publisher} {Oxford
    University Press},\ \bibinfo {address} {Oxford},\ \bibinfo {year}
  {2012})\BibitemShut {NoStop}%
\bibitem [{\citenamefont {Greiner}\ \emph {et~al.}(2002)\citenamefont
    {Greiner}, \citenamefont {Mandel}, \citenamefont {Esslinger},
    \citenamefont {H\"ansch},\ and\ \citenamefont
    {Bloch}}]{greiner2002}%
  \BibitemOpen \bibfield {author} {\bibinfo {author} {\bibfnamefont
      {M.}~\bibnamefont {Greiner}}, \bibinfo {author} {\bibfnamefont
      {O.}~\bibnamefont {Mandel}}, \bibinfo {author} {\bibfnamefont
      {T.}~\bibnamefont {Esslinger}}, \bibinfo {author} {\bibfnamefont
      {T.~W.}\ \bibnamefont {H\"ansch}}, \ and\ \bibinfo {author}
    {\bibfnamefont {I.}~\bibnamefont {Bloch}},\ }\href {\doibase
    10.1038/415039a} {\bibfield {journal} {\bibinfo {journal}
      {Nature}\ }\textbf {\bibinfo {volume} {415}},\ \bibinfo {pages}
    {39} (\bibinfo {year} {2002})}\BibitemShut {NoStop}%
\bibitem [{\citenamefont {Strohmaier}\ \emph
    {et~al.}(2010)\citenamefont {Strohmaier}, \citenamefont {Greif},
    \citenamefont {J\"ordens}, \citenamefont {Tarruell}, \citenamefont
    {Moritz}, \citenamefont {Esslinger}, \citenamefont {Sensarma},
    \citenamefont {Pekker}, \citenamefont {Altman},\ and\
    \citenamefont {Demler}}]{strohmaier2010}%
  \BibitemOpen \bibfield {author} {\bibinfo {author} {\bibfnamefont
      {N.}~\bibnamefont {Strohmaier}}, \bibinfo {author}
    {\bibfnamefont {D.}~\bibnamefont {Greif}}, \bibinfo {author}
    {\bibfnamefont {R.}~\bibnamefont {J\"ordens}}, \bibinfo {author}
    {\bibfnamefont {L.}~\bibnamefont {Tarruell}}, \bibinfo {author}
    {\bibfnamefont {H.}~\bibnamefont {Moritz}}, \bibinfo {author}
    {\bibfnamefont {T.}~\bibnamefont {Esslinger}}, \bibinfo {author}
    {\bibfnamefont {R.}~\bibnamefont {Sensarma}}, \bibinfo {author}
    {\bibfnamefont {D.}~\bibnamefont {Pekker}}, \bibinfo {author}
    {\bibfnamefont {E.}~\bibnamefont {Altman}}, \ and\ \bibinfo
    {author} {\bibfnamefont {E.}~\bibnamefont {Demler}},\ }\href
  {\doibase 10.1103/PhysRevLett.104.080401} {\bibfield {journal}
    {\bibinfo {journal} {Phys. Rev. Lett.}\ }\textbf {\bibinfo
      {volume} {104}},\ \bibinfo {pages} {080401} (\bibinfo {year}
    {2010})}
  \BibitemShut {NoStop}%
\bibitem [{\citenamefont {Iwai}\ \emph {et~al.}(2003)\citenamefont
    {Iwai}, \citenamefont {Ono}, \citenamefont {Maeda}, \citenamefont
    {Matsuzaki}, \citenamefont {Kishida}, \citenamefont {Okamoto},\
    and\ \citenamefont {Tokura}}]{iwai2003}%
  \BibitemOpen \bibfield {author} {\bibinfo {author} {\bibfnamefont
      {S.}~\bibnamefont {Iwai}}, \bibinfo {author} {\bibfnamefont
      {M.}~\bibnamefont {Ono}}, \bibinfo {author} {\bibfnamefont
      {A.}~\bibnamefont {Maeda}}, \bibinfo {author} {\bibfnamefont
      {H.}~\bibnamefont {Matsuzaki}}, \bibinfo {author} {\bibfnamefont
      {H.}~\bibnamefont {Kishida}}, \bibinfo {author} {\bibfnamefont
      {H.}~\bibnamefont {Okamoto}}, \ and\ \bibinfo {author}
    {\bibfnamefont {Y.}~\bibnamefont {Tokura}},\ }\href {\doibase
    10.1103/PhysRevLett.91.057401} {\bibfield {journal} {\bibinfo
      {journal} {Phys. Rev. Lett.}\ }\textbf {\bibinfo {volume}
      {91}},\ \bibinfo {pages} {057401} (\bibinfo {year}
    {2003})}\BibitemShut {NoStop}%
\bibitem [{\citenamefont {Perfetti}\ \emph {et~al.}(2006)\citenamefont
    {Perfetti}, \citenamefont {Loukakos}, \citenamefont {Lisowski},
    \citenamefont {Bovensiepen}, \citenamefont {Berger}, \citenamefont
    {Biermann}, \citenamefont {Cornaglia}, \citenamefont {Georges},\
    and\ \citenamefont {Wolf}}]{perfetti2006}%
  \BibitemOpen \bibfield {author} {\bibinfo {author} {\bibfnamefont
      {L.}~\bibnamefont {Perfetti}}, \bibinfo {author} {\bibfnamefont
      {P.~A.}\ \bibnamefont {Loukakos}}, \bibinfo {author}
    {\bibfnamefont {M.}~\bibnamefont {Lisowski}}, \bibinfo {author}
    {\bibfnamefont {U.}~\bibnamefont {Bovensiepen}}, \bibinfo {author}
    {\bibfnamefont {H.}~\bibnamefont {Berger}}, \bibinfo {author}
    {\bibfnamefont {S.}~\bibnamefont {Biermann}}, \bibinfo {author}
    {\bibfnamefont {P.~S.}\ \bibnamefont {Cornaglia}}, \bibinfo
    {author} {\bibfnamefont {A.}~\bibnamefont {Georges}}, \ and\
    \bibinfo {author} {\bibfnamefont {M.}~\bibnamefont {Wolf}},\
  }\href {\doibase 10.1103/PhysRevLett.97.067402} {\bibfield {journal}
    {\bibinfo {journal} {Phys. Rev. Lett.}\ }\textbf {\bibinfo
      {volume} {97}},\ \bibinfo {pages} {067402} (\bibinfo {year}
    {2006})}\BibitemShut {NoStop}%
\bibitem [{\citenamefont {Wall}\ \emph {et~al.}(2011)\citenamefont
    {Wall}, \citenamefont {Brida}, \citenamefont {Clark},
    \citenamefont {Ehrke}, \citenamefont {Jaksch}, \citenamefont
    {Ardavan}, \citenamefont {Bonora}, \citenamefont {Uemura},
    \citenamefont {Takahashi}, \citenamefont {Hasegawa}, \citenamefont
    {Okamoto}, \citenamefont {Cerullo},\ and\ \citenamefont
    {Cavalleri}}]{wall2011}%
  \BibitemOpen \bibfield {author} {\bibinfo {author} {\bibfnamefont
      {S.}~\bibnamefont {Wall}}, \bibinfo {author} {\bibfnamefont
      {D.}~\bibnamefont {Brida}}, \bibinfo {author} {\bibfnamefont
      {S.~R.}\ \bibnamefont {Clark}}, \bibinfo {author} {\bibfnamefont
      {H.~P.}\ \bibnamefont {Ehrke}}, \bibinfo {author} {\bibfnamefont
      {D.}~\bibnamefont {Jaksch}}, \bibinfo {author} {\bibfnamefont
      {A.}~\bibnamefont {Ardavan}}, \bibinfo {author} {\bibfnamefont
      {S.}~\bibnamefont {Bonora}}, \bibinfo {author} {\bibfnamefont
      {H.}~\bibnamefont {Uemura}}, \bibinfo {author} {\bibfnamefont
      {Y.}~\bibnamefont {Takahashi}}, \bibinfo {author} {\bibfnamefont
      {T.}~\bibnamefont {Hasegawa}}, \bibinfo {author} {\bibfnamefont
      {H.}~\bibnamefont {Okamoto}}, \bibinfo {author} {\bibfnamefont
      {G.}~\bibnamefont {Cerullo}}, \ and\ \bibinfo {author}
    {\bibfnamefont {A.}~\bibnamefont {Cavalleri}},\ }\href {\doibase
    10.1038/nphys1831} {\bibfield {journal} {\bibinfo {journal}
      {Nat. Phys.}\ }\textbf {\bibinfo {volume} {7}},\ \bibinfo
    {pages} {114} (\bibinfo {year} {2011})}\BibitemShut {NoStop}%
\bibitem [{\citenamefont {Deutsch}(1991)}]{deutsch1991}%
  \BibitemOpen \bibfield {author} {\bibinfo {author} {\bibfnamefont
      {J.~M.}\ \bibnamefont {Deutsch}},\ }\href {\doibase
    10.1103/PhysRevA.43.2046} {\bibfield {journal} {\bibinfo {journal}
      {Phys. Rev. A}\ }\textbf {\bibinfo {volume} {43}},\ \bibinfo
    {pages} {2046} (\bibinfo {year} {1991})}\BibitemShut {NoStop}%
\bibitem [{\citenamefont {Srednicki}(1994)}]{srednicki1994}%
  \BibitemOpen \bibfield {author} {\bibinfo {author} {\bibfnamefont
      {M.}~\bibnamefont {Srednicki}},\ }\href {\doibase
    10.1103/PhysRevE.50.888} {\bibfield {journal} {\bibinfo {journal}
      {Phys. Rev. E}\ }\textbf {\bibinfo {volume} {50}},\ \bibinfo
    {pages} {888} (\bibinfo {year} {1994})}\BibitemShut {NoStop}%
\bibitem [{\citenamefont {Rigol}\ \emph {et~al.}(2008)\citenamefont
    {Rigol}, \citenamefont {Dunjko},\ and\ \citenamefont
    {Olshanii}}]{rigol2008}%
  \BibitemOpen \bibfield {author} {\bibinfo {author} {\bibfnamefont
      {M.}~\bibnamefont {Rigol}}, \bibinfo {author} {\bibfnamefont
      {V.}~\bibnamefont {Dunjko}}, \ and\ \bibinfo {author}
    {\bibfnamefont {M.}~\bibnamefont {Olshanii}},\ }\href {\doibase
    10.1038/nature06838} {\bibfield {journal} {\bibinfo {journal}
      {Nature}\ }\textbf {\bibinfo {volume} {452}},\ \bibinfo {pages}
    {854} (\bibinfo {year} {2008})}\BibitemShut {NoStop}%
\bibitem [{\citenamefont {Dziarmaga}(2010)}]{dziarmaga2010}%
  \BibitemOpen \bibfield {author} {\bibinfo {author} {\bibfnamefont
      {J.}~\bibnamefont {Dziarmaga}},\ }\href {\doibase
    10.1080/00018732.2010.514702} {\bibfield {journal} {\bibinfo
      {journal} {Adv. Phys.}\ }\textbf {\bibinfo {volume} {59}},\
    \bibinfo {pages} {1063} (\bibinfo {year} {2010})}
  \BibitemShut {NoStop}%
\bibitem [{\citenamefont {Polkovnikov}\ \emph
    {et~al.}(2011)\citenamefont {Polkovnikov}, \citenamefont
    {Sengupta}, \citenamefont {Silva},\ and\ \citenamefont
    {Vengalattore}}]{polkovnikov2011}%
  \BibitemOpen \bibfield {author} {\bibinfo {author} {\bibfnamefont
      {A.}~\bibnamefont {Polkovnikov}}, \bibinfo {author}
    {\bibfnamefont {K.}~\bibnamefont {Sengupta}}, \bibinfo {author}
    {\bibfnamefont {A.}~\bibnamefont {Silva}}, \ and\ \bibinfo
    {author} {\bibfnamefont {M.}~\bibnamefont {Vengalattore}},\ }\href
  {\doibase 10.1103/RevModPhys.83.863} {\bibfield {journal} {\bibinfo
      {journal} {Rev. Mod. Phys.}\ }\textbf {\bibinfo {volume} {83}},\
    \bibinfo {pages} {863} (\bibinfo {year} {2011})}
  \BibitemShut {NoStop}%
\bibitem [{\citenamefont {Metzner}\ and\ \citenamefont
    {Vollhardt}(1989)}]{metzner1989}%
  \BibitemOpen \bibfield {author} {\bibinfo {author} {\bibfnamefont
      {W.}~\bibnamefont {Metzner}}\ and\ \bibinfo {author}
    {\bibfnamefont {D.}~\bibnamefont {Vollhardt}},\ }\href {\doibase
    10.1103/PhysRevLett.62.324} {\bibfield {journal} {\bibinfo
      {journal} {Phys. Rev. Lett.}\ }\textbf {\bibinfo {volume}
      {62}},\ \bibinfo {pages} {324} (\bibinfo {year}
    {1989})}\BibitemShut {NoStop}%
\bibitem [{\citenamefont {Georges}\ \emph {et~al.}(1996)\citenamefont
    {Georges}, \citenamefont {Kotliar}, \citenamefont {Krauth},\ and\
    \citenamefont {Rozenberg}}]{georges1996}%
  \BibitemOpen \bibfield {author} {\bibinfo {author} {\bibfnamefont
      {A.}~\bibnamefont {Georges}}, \bibinfo {author} {\bibfnamefont
      {G.}~\bibnamefont {Kotliar}}, \bibinfo {author} {\bibfnamefont
      {W.}~\bibnamefont {Krauth}}, \ and\ \bibinfo {author}
    {\bibfnamefont {M.~J.}\ \bibnamefont {Rozenberg}},\ }\href
  {\doibase 10.1103/RevModPhys.68.13} {\bibfield {journal} {\bibinfo
      {journal} {Rev.  Mod. Phys.}\ }\textbf {\bibinfo {volume}
      {68}},\ \bibinfo {pages} {13} (\bibinfo {year}
    {1996})}\BibitemShut {NoStop}%
\bibitem [{\citenamefont {Schmidt}\ and\ \citenamefont
    {Monien}(2002)}]{schmidt2002}%
  \BibitemOpen \bibfield {author} {\bibinfo {author} {\bibfnamefont
      {P.}~\bibnamefont {Schmidt}}\ and\ \bibinfo {author}
    {\bibfnamefont {H.}~\bibnamefont {Monien}},\ }\href@noop {}
  \Eprint {http://arxiv.org/abs/cond-mat/0202046} {cond-mat/0202046},\
  {(\bibinfo {year} {2002})} \BibitemShut {NoStop}%
\bibitem [{\citenamefont {Freericks}\ \emph
    {et~al.}(2006)\citenamefont {Freericks}, \citenamefont
    {Turkowski},\ and\ \citenamefont {Zlati\'c}}]{freericks2006}%
  \BibitemOpen \bibfield {author} {\bibinfo {author} {\bibfnamefont
      {J.~K.}\ \bibnamefont {Freericks}}, \bibinfo {author}
    {\bibfnamefont {V.~M.}\ \bibnamefont {Turkowski}}, \ and\ \bibinfo
    {author} {\bibfnamefont {V.}~\bibnamefont {Zlati\'c}},\ }\href
  {\doibase 10.1103/PhysRevLett.97.266408} {\bibfield {journal}
    {\bibinfo {journal} {Phys. Rev. Lett.}\ }\textbf {\bibinfo
      {volume} {97}},\ \bibinfo {eid} {266408} (\bibinfo {year}
    {2006})}
  \BibitemShut {NoStop}%
\bibitem [{\citenamefont {Aoki}\ \emph {et~al.}(2014)\citenamefont
    {Aoki}, \citenamefont {Tsuji}, \citenamefont {Eckstein},
    \citenamefont {Kollar}, \citenamefont {Oka},\ and\ \citenamefont
    {Werner}}]{aoki2013}%
  \BibitemOpen \bibfield {author} {\bibinfo {author} {\bibfnamefont
      {H.}~\bibnamefont {Aoki}}, \bibinfo {author} {\bibfnamefont
      {N.}~\bibnamefont {Tsuji}}, \bibinfo {author} {\bibfnamefont
      {M.}~\bibnamefont {Eckstein}}, \bibinfo {author} {\bibfnamefont
      {M.}~\bibnamefont {Kollar}}, \bibinfo {author} {\bibfnamefont
      {T.}~\bibnamefont {Oka}}, \ and\ \bibinfo {author}
    {\bibfnamefont {P.}~\bibnamefont {Werner}},\ }\href {\doibase
    10.1103/RevModPhys.86.779} {\bibfield {journal} {\bibinfo
      {journal} {Rev.  Mod. Phys.}\ }\textbf {\bibinfo {volume}
      {86}},\ \bibinfo {pages} {779} (\bibinfo {year}
    {2014})}
  \BibitemShut {NoStop}%
\bibitem [{\citenamefont {Moeckel}\ and\ \citenamefont
    {Kehrein}(2008)}]{moeckel2008}%
  \BibitemOpen \bibfield {author} {\bibinfo {author} {\bibfnamefont
      {M.}~\bibnamefont {Moeckel}}\ and\ \bibinfo {author}
    {\bibfnamefont {S.}~\bibnamefont {Kehrein}},\ }\href {\doibase
    10.1103/PhysRevLett.100.175702} {\bibfield {journal} {\bibinfo
      {journal} {Phys. Rev. Lett.}\ }\textbf {\bibinfo {volume}
      {100}},\ \bibinfo {eid} {175702} (\bibinfo {year}
    {2008})}
  \BibitemShut {NoStop}%
\bibitem [{\citenamefont {Eckstein}\ \emph {et~al.}(2009)\citenamefont
    {Eckstein}, \citenamefont {Kollar},\ and\ \citenamefont
    {Werner}}]{eckstein2009b}%
  \BibitemOpen \bibfield {author} {\bibinfo {author} {\bibfnamefont
      {M.}~\bibnamefont {Eckstein}}, \bibinfo {author} {\bibfnamefont
      {M.}~\bibnamefont {Kollar}}, \ and\ \bibinfo {author}
    {\bibfnamefont {P.}~\bibnamefont {Werner}},\ }\href {\doibase
    10.1103/PhysRevLett.103.056403} {\bibfield {journal} {\bibinfo
      {journal} {Phys. Rev. Lett.}\ }\textbf {\bibinfo {volume}
      {103}},\ \bibinfo {eid} {056403} (\bibinfo {year}
    {2009})}
  \BibitemShut {NoStop}%
\bibitem [{\citenamefont {Schir\'o}\ and\ \citenamefont
    {Fabrizio}(2010)}]{schiro2010b}%
  \BibitemOpen \bibfield {author} {\bibinfo {author} {\bibfnamefont
      {M.}~\bibnamefont {Schir\'o}}\ and\ \bibinfo {author}
    {\bibfnamefont {M.}~\bibnamefont {Fabrizio}},\ }\href {\doibase
    10.1103/PhysRevLett.105.076401} {\bibfield {journal} {\bibinfo
      {journal} {Phys. Rev. Lett.}\ }\textbf {\bibinfo {volume}
      {105}},\ \bibinfo {eid} {076401} (\bibinfo {year}
    {2010})}
  \BibitemShut {NoStop}%
\bibitem [{\citenamefont {Schir\'o}\ and\ \citenamefont
    {Fabrizio}(2011)}]{schiro2011}%
  \BibitemOpen \bibfield {author} {\bibinfo {author} {\bibfnamefont
      {M.}~\bibnamefont {Schir\'o}}\ and\ \bibinfo {author}
    {\bibfnamefont {M.}~\bibnamefont {Fabrizio}},\ }\href {\doibase
    10.1103/PhysRevB.83.165105} {\bibfield {journal} {\bibinfo
      {journal} {Phys. Rev. B}\ }\textbf {\bibinfo {volume} {83}},\
    \bibinfo {eid} {165105} (\bibinfo {year} {2011})}
  \BibitemShut {NoStop}%
\bibitem [{\citenamefont {Sandri}\ \emph {et~al.}(2012)\citenamefont
    {Sandri}, \citenamefont {Schir\'o},\ and\ \citenamefont
    {Fabrizio}}]{sandri2012}%
  \BibitemOpen \bibfield {author} {\bibinfo {author} {\bibfnamefont
      {M.}~\bibnamefont {Sandri}}, \bibinfo {author} {\bibfnamefont
      {M.}~\bibnamefont {Schir\'o}}, \ and\ \bibinfo {author}
    {\bibfnamefont {M.}~\bibnamefont {Fabrizio}},\ }\href {\doibase
    10.1103/PhysRevB.86.075122} {\bibfield {journal} {\bibinfo
      {journal} {Phys. Rev. B}\ }\textbf {\bibinfo {volume} {86}},\
    \bibinfo {eid} {075122} (\bibinfo {year}
    {2012})}
  \BibitemShut {NoStop}%
\bibitem [{\citenamefont {Hamerla}\ and\ \citenamefont
    {Uhrig}(2013)}]{hamerla2013}%
  \BibitemOpen \bibfield {author} {\bibinfo {author} {\bibfnamefont
      {S.~A.}\ \bibnamefont {Hamerla}}\ and\ \bibinfo {author}
    {\bibfnamefont {G.~S.}\ \bibnamefont {Uhrig}},\ }\href {\doibase
    10.1103/PhysRevB.87.064304} {\bibfield {journal} {\bibinfo
      {journal} {Phys. Rev. B}\ }\textbf {\bibinfo {volume} {87}},\
    \bibinfo {eid} {064304} (\bibinfo {year} {2013})}
  \BibitemShut {NoStop}%
\bibitem [{\citenamefont {Hamerla}\ and\ \citenamefont
    {Uhrig}(2014)}]{hamerla2014}%
  \BibitemOpen \bibfield {author} {\bibinfo {author} {\bibfnamefont
      {S.~A.}\ \bibnamefont {Hamerla}}\ and\ \bibinfo {author}
    {\bibfnamefont {G.~S.}\ \bibnamefont {Uhrig}},\ }\href {\doibase
    10.1103/PhysRevB.89.104301} {\bibfield {journal} {\bibinfo
      {journal} {Phys. Rev. B}\ }\textbf {\bibinfo {volume} {89}},\
    \bibinfo {eid} {104301} (\bibinfo {year} {2014})}
  \BibitemShut {NoStop}%
\bibitem [{\citenamefont {Werner}\ \emph {et~al.}(2009)\citenamefont
    {Werner}, \citenamefont {Oka},\ and\ \citenamefont
    {Millis}}]{werner2009}%
  \BibitemOpen \bibfield {author} {\bibinfo {author} {\bibfnamefont
      {P.}~\bibnamefont {Werner}}, \bibinfo {author} {\bibfnamefont
      {T.}~\bibnamefont {Oka}}, \ and\ \bibinfo {author}
    {\bibfnamefont {A.~J.}\ \bibnamefont {Millis}},\ }\href {\doibase
    10.1103/PhysRevB.79.035320} {\bibfield {journal} {\bibinfo
      {journal} {Phys. Rev. B}\ }\textbf {\bibinfo {volume} {79}},\
    \bibinfo {eid} {035320} (\bibinfo {year}
    {2009})}
  \BibitemShut {NoStop}%
\bibitem [{\citenamefont {Gull}\ \emph {et~al.}(2011)\citenamefont
    {Gull}, \citenamefont {Millis}, \citenamefont {Lichtenstein},
    \citenamefont {Rubtsov}, \citenamefont {Troyer},\ and\
    \citenamefont {Werner}}]{gull2011}%
  \BibitemOpen \bibfield {author} {\bibinfo {author} {\bibfnamefont
      {E.}~\bibnamefont {Gull}}, \bibinfo {author} {\bibfnamefont
      {A.~J.}\ \bibnamefont {Millis}}, \bibinfo {author}
    {\bibfnamefont {A.~I.}\ \bibnamefont {Lichtenstein}}, \bibinfo
    {author} {\bibfnamefont {A.~N.}\ \bibnamefont {Rubtsov}}, \bibinfo
    {author} {\bibfnamefont {M.}~\bibnamefont {Troyer}}, \ and\
    \bibinfo {author} {\bibfnamefont {P.}~\bibnamefont {Werner}},\
  }\href {\doibase 10.1103/RevModPhys.83.349} {\bibfield {journal}
    {\bibinfo {journal} {Rev.  Mod. Phys.}\ }\textbf {\bibinfo
      {volume} {83}},\ \bibinfo {pages} {349} (\bibinfo {year}
    {2011})}
  \BibitemShut {NoStop}%
\bibitem [{\citenamefont {Eckstein}\ and\ \citenamefont
    {Werner}(2010)}]{eckstein2010}%
  \BibitemOpen \bibfield {author} {\bibinfo {author} {\bibfnamefont
      {M.}~\bibnamefont {Eckstein}}\ and\ \bibinfo {author}
    {\bibfnamefont {P.}~\bibnamefont {Werner}},\ }\href {\doibase
    10.1103/PhysRevB.82.115115} {\bibfield {journal} {\bibinfo
      {journal} {Phys. Rev. B}\ }\textbf {\bibinfo {volume} {82}},\
    \bibinfo {eid} {115115} (\bibinfo {year} {2010})}
  \BibitemShut {NoStop}%
\bibitem [{\citenamefont {Eckstein}\ and\ \citenamefont
    {Werner}(2011)}]{eckstein2011b}%
  \BibitemOpen \bibfield {author} {\bibinfo {author} {\bibfnamefont
      {M.}~\bibnamefont {Eckstein}}\ and\ \bibinfo {author}
    {\bibfnamefont {P.}~\bibnamefont {Werner}},\ }\href {\doibase
    10.1103/PhysRevLett.107.186406} {\bibfield {journal} {\bibinfo
      {journal} {Phys. Rev. Lett.}\ }\textbf {\bibinfo {volume}
      {107}},\ \bibinfo {eid} {186406} (\bibinfo {year}
    {2011})}
  \BibitemShut {NoStop}%
\bibitem [{\citenamefont {Tsuji}\ \emph {et~al.}(2013)\citenamefont
    {Tsuji}, \citenamefont {Eckstein},\ and\ \citenamefont
    {Werner}}]{tsuji2013b}%
  \BibitemOpen \bibfield {author} {\bibinfo {author} {\bibfnamefont
      {N.}~\bibnamefont {Tsuji}}, \bibinfo {author} {\bibfnamefont
      {M.}~\bibnamefont {Eckstein}}, \ and\ \bibinfo {author}
    {\bibfnamefont {P.}~\bibnamefont {Werner}},\ }\href {\doibase
    10.1103/PhysRevLett.110.136404} {\bibfield {journal} {\bibinfo
      {journal} {Phys. Rev. Lett.}\ }\textbf {\bibinfo {volume}
      {110}},\ \bibinfo {eid} {136404} (\bibinfo {year}
    {2013})}
  \BibitemShut {NoStop}%
\bibitem [{\citenamefont {Caffarel}\ and\ \citenamefont
    {Krauth}(1994)}]{caffarel1994}%
  \BibitemOpen \bibfield {author} {\bibinfo {author} {\bibfnamefont
      {M.}~\bibnamefont {Caffarel}}\ and\ \bibinfo {author}
    {\bibfnamefont {W.}~\bibnamefont {Krauth}},\ }\href {\doibase
    10.1103/PhysRevLett.72.1545} {\bibfield {journal} {\bibinfo
      {journal} {Phys. Rev. Lett.}\ }\textbf {\bibinfo {volume}
      {72}},\ \bibinfo {pages} {1545} (\bibinfo {year}
    {1994})}\BibitemShut {NoStop}%
\bibitem [{\citenamefont {Gramsch}\ \emph {et~al.}(2013)\citenamefont
    {Gramsch}, \citenamefont {Balzer}, \citenamefont {Eckstein},\ and\
    \citenamefont {Kollar}}]{gramsch2013}%
  \BibitemOpen \bibfield {author} {\bibinfo {author} {\bibfnamefont
      {C.}~\bibnamefont {Gramsch}}, \bibinfo {author} {\bibfnamefont
      {K.}~\bibnamefont {Balzer}}, \bibinfo {author} {\bibfnamefont
      {M.}~\bibnamefont {Eckstein}}, \ and\ \bibinfo {author}
    {\bibfnamefont {M.}~\bibnamefont {Kollar}},\ }\href {\doibase
    10.1103/PhysRevB.88.235106} {\bibfield {journal} {\bibinfo
      {journal} {Phys. Rev. B}\ }\textbf {\bibinfo {volume} {88}},\
    \bibinfo {pages} {235106} (\bibinfo {year} {2013})}
  \BibitemShut {NoStop}%
\bibitem [{\citenamefont {Balzer}\ \emph {et~al.}(2015)\citenamefont
    {Balzer}, \citenamefont {Li}, \citenamefont {Vendrell},\ and\
    \citenamefont {Eckstein}}]{balzer2015}%
  \BibitemOpen \bibfield {author} {\bibinfo {author} {\bibfnamefont
      {K.}~\bibnamefont {Balzer}}, \bibinfo {author} {\bibfnamefont
      {Z.}~\bibnamefont {Li}}, \bibinfo {author} {\bibfnamefont
      {O.}~\bibnamefont {Vendrell}}, \ and\ \bibinfo {author}
    {\bibfnamefont {M.}~\bibnamefont {Eckstein}},\ }\href {\doibase
    10.1103/PhysRevB.91.045136} {\bibfield {journal} {\bibinfo
      {journal} {Phys.  Rev. B}\ }\textbf {\bibinfo {volume} {91}},\
    \bibinfo {eid} {045136} (\bibinfo {year}
    {2015})}
  \BibitemShut {NoStop}%
\bibitem [{\citenamefont {Wolf}\ \emph {et~al.}(2014)\citenamefont
    {Wolf}, \citenamefont {McCulloch},\ and\ \citenamefont
    {Schollw\"ock}}]{wolf2014}%
  \BibitemOpen \bibfield {author} {\bibinfo {author} {\bibfnamefont
      {F.~A.}\ \bibnamefont {Wolf}}, \bibinfo {author} {\bibfnamefont
      {I.~P.}\ \bibnamefont {McCulloch}}, \ and\ \bibinfo {author}
    {\bibfnamefont {U.}~\bibnamefont {Schollw\"ock}},\ }\href
  {\doibase 10.1103/PhysRevB.90.235131} {\bibfield {journal} {\bibinfo
      {journal} {Phys. Rev. B}\ }\textbf {\bibinfo {volume} {90}},\
    \bibinfo {eid} {235131} (\bibinfo {year}
    {2014})}
  \BibitemShut {NoStop}%
\bibitem [{\citenamefont
    {Potthoff}(2003{\natexlab{a}})}]{potthoff2003}%
  \BibitemOpen \bibfield {author} {\bibinfo {author} {\bibfnamefont
      {M.}~\bibnamefont {Potthoff}},\ }\href {\doibase
    10.1140/epjb/e2003-00121-8} {\bibfield {journal} {\bibinfo
      {journal} {Eur. Phys. J. B}\ }\textbf {\bibinfo {volume} {32}},\
    \bibinfo {pages} {429} (\bibinfo {year} {2003}{\natexlab{a}})}
  \BibitemShut {NoStop}%
\bibitem [{\citenamefont {Hofmann}\ \emph {et~al.}(2013)\citenamefont
    {Hofmann}, \citenamefont {Eckstein}, \citenamefont {Arrigoni},\
    and\ \citenamefont {Potthoff}}]{hofmann2013}%
  \BibitemOpen \bibfield {author} {\bibinfo {author} {\bibfnamefont
      {F.}~\bibnamefont {Hofmann}}, \bibinfo {author} {\bibfnamefont
      {M.}~\bibnamefont {Eckstein}}, \bibinfo {author} {\bibfnamefont
      {E.}~\bibnamefont {Arrigoni}}, \ and\ \bibinfo {author}
    {\bibfnamefont {M.}~\bibnamefont {Potthoff}},\ }\href {\doibase
    10.1103/PhysRevB.88.165124} {\bibfield {journal} {\bibinfo
      {journal} {Phys. Rev. B}\ }\textbf {\bibinfo {volume} {88}},\
    \bibinfo {eid} {165124} (\bibinfo {year}
    {2013})}
  \BibitemShut {NoStop}%
\bibitem [{\citenamefont {Potthoff}(2012)}]{potthoff2012}%
  \BibitemOpen \bibfield {author} {\bibinfo {author} {\bibfnamefont
      {M.}~\bibnamefont {Potthoff}},\ }in\ \href {\doibase
    10.1007/978-3-642-21831-6_10} {\emph {\bibinfo {booktitle}
      {Strongly Correlated Systems: Theoretical Methods}}},\ \bibinfo
  {series} {Springer Series in Solid-State Sciences}, Vol.\ \bibinfo
  {volume} {171},\ \bibinfo {editor} {edited by\ \bibinfo {editor}
    {\bibfnamefont {A.}~\bibnamefont {Avella}}\ and\ \bibinfo {editor}
    {\bibfnamefont {F.}~\bibnamefont {Mancini}}}\ (\bibinfo
  {publisher} {Springer},\ \bibinfo {address} {Berlin},\ \bibinfo
  {year} {2012})
  \BibitemShut {NoStop}%
\bibitem [{\citenamefont {Potthoff}(2014)}]{potthoff2014}%
  \BibitemOpen \bibfield {author} {\bibinfo {author} {\bibfnamefont
      {M.}~\bibnamefont {Potthoff}},\ }in\ \href@noop {} {\emph
    {\bibinfo {booktitle} {DMFT at 25: Infinite Dimensions}}},\
  \bibinfo {series} {Modeling and Simulation}, Vol.~\bibinfo {volume}
  {4},\ \bibinfo {editor} {edited by\ \bibinfo {editor} {\bibfnamefont
      {D.~V.}\ \bibnamefont {Eva~Pavarini}, \bibfnamefont
      {Erik~Koch}}\ and\ \bibinfo {editor} {\bibfnamefont
      {A.}~\bibnamefont {Lichtenstein}}}\ (\bibinfo {publisher}
  {Forschungszentrum J\"ulich},\ \bibinfo {year}
  {2014})\ 
  \BibitemShut {NoStop}%
\bibitem [{\citenamefont
    {Potthoff}(2003{\natexlab{b}})}]{potthoff2003b}%
  \BibitemOpen \bibfield {author} {\bibinfo {author} {\bibfnamefont
      {M.}~\bibnamefont {Potthoff}},\ }\href {\doibase
    10.1140/epjb/e2003-00352-7} {\bibfield {journal} {\bibinfo
      {journal} {Eur. Phys. J. B}\ }\textbf {\bibinfo {volume} {36}},\
    \bibinfo {pages} {335} (\bibinfo {year} {2003}{\natexlab{b}})}
  \BibitemShut {NoStop}%
\bibitem [{\citenamefont
    {Po\v{z}gaj\v{c}i\'{c}}(2004)}]{pozgajcic2004}%
  \BibitemOpen \bibfield {author} {\bibinfo {author} {\bibfnamefont
      {K.}~\bibnamefont {Po\v{z}gaj\v{c}i\'{c}}},\ }\href@noop {}
  \Eprint {http://arxiv.org/abs/cond-mat/0407172} {cond-mat/0407172},\
  {(\bibinfo {year} {2004})} \BibitemShut {NoStop}%
\bibitem [{\citenamefont {Koga}\ \emph {et~al.}(2005)\citenamefont
    {Koga}, \citenamefont {Inaba},\ and\ \citenamefont
    {Kawakami}}]{koga2005}%
  \BibitemOpen \bibfield {author} {\bibinfo {author} {\bibfnamefont
      {A.}~\bibnamefont {Koga}}, \bibinfo {author} {\bibfnamefont
      {K.}~\bibnamefont {Inaba}}, \ and\ \bibinfo {author}
    {\bibfnamefont {N.}~\bibnamefont {Kawakami}},\ }\href {\doibase
    10.1143/PTPS.160.253} {\bibfield {journal} {\bibinfo {journal}
      {Progress of Theoretical Physics Supplement}\ }\textbf {\bibinfo
      {volume} {160}},\ \bibinfo {pages} {253} (\bibinfo {year}
    {2005})}
  \BibitemShut {NoStop}%
\bibitem [{\citenamefont {Inaba}\ \emph {et~al.}(2005)\citenamefont
    {Inaba}, \citenamefont {Koga}, \citenamefont {Suga},\ and\
    \citenamefont {Kawakami}}]{inaba2005b}%
  \BibitemOpen \bibfield {author} {\bibinfo {author} {\bibfnamefont
      {K.}~\bibnamefont {Inaba}}, \bibinfo {author} {\bibfnamefont
      {A.}~\bibnamefont {Koga}}, \bibinfo {author} {\bibfnamefont
      {S.~I.}\ \bibnamefont {Suga}}, \ and\ \bibinfo {author}
    {\bibfnamefont {N.}~\bibnamefont {Kawakami}},\ }\href {\doibase
    10.1103/PhysRevB.72.085112} {\bibfield {journal} {\bibinfo
      {journal} {Phys. Rev. B}\ }\textbf {\bibinfo {volume} {72}},\
    \bibinfo {eid} {085112} (\bibinfo {year} {2005})}
  \BibitemShut {NoStop}%
\bibitem [{\citenamefont {Eckstein}\ \emph {et~al.}(2007)\citenamefont
    {Eckstein}, \citenamefont {Kollar}, \citenamefont {Potthoff},\
    and\ \citenamefont {Vollhardt}}]{eckstein2007}%
  \BibitemOpen \bibfield {author} {\bibinfo {author} {\bibfnamefont
      {M.}~\bibnamefont {Eckstein}}, \bibinfo {author} {\bibfnamefont
      {M.}~\bibnamefont {Kollar}}, \bibinfo {author} {\bibfnamefont
      {M.}~\bibnamefont {Potthoff}}, \ and\ \bibinfo {author}
    {\bibfnamefont {D.}~\bibnamefont {Vollhardt}},\ }\href {\doibase
    10.1103/PhysRevB.75.125103} {\bibfield {journal} {\bibinfo
      {journal} {Phys. Rev. B}\ }\textbf {\bibinfo {volume} {75}},\
    \bibinfo {eid} {125103} (\bibinfo {year} {2007})}
  \BibitemShut {NoStop}%
\bibitem [{\citenamefont {Hofmann}\ \emph {et~al.}(2016)\citenamefont
    {Hofmann}, \citenamefont {Eckstein},\ and\ \citenamefont
    {Potthoff}}]{hofmann2015}%
  \BibitemOpen \bibfield {author} {\bibinfo {author} {\bibfnamefont
      {F.}~\bibnamefont {Hofmann}}, \bibinfo {author} {\bibfnamefont
      {M.}~\bibnamefont {Eckstein}}, \ and\ \bibinfo {author}
    {\bibfnamefont {M.}~\bibnamefont {Potthoff}},\ }\href {\doibase
    10.1088/1742-6596/696/1/012002} {\bibfield {journal} {\bibinfo
      {journal} {Journal of Physics: Conference Series}\ }\textbf
    {\bibinfo {volume} {696}},\ \bibinfo {pages} {012002} (\bibinfo
    {year} {2016})}
  \BibitemShut {NoStop}%
\bibitem [{\citenamefont {Wagner}(1991)}]{wagner1991}%
  \BibitemOpen \bibfield {author} {\bibinfo {author} {\bibfnamefont
      {M.}~\bibnamefont {Wagner}},\ }\href {\doibase
    10.1103/PhysRevB.44.6104} {\bibfield {journal} {\bibinfo {journal}
      {Phys. Rev. B}\ }\textbf {\bibinfo {volume} {44}},\ \bibinfo
    {pages} {6104} (\bibinfo {year} {1991})}\BibitemShut {NoStop}%
\bibitem [{\citenamefont {van Leeuwen}\ \emph
    {et~al.}(2006)\citenamefont {van Leeuwen}, \citenamefont {Dahlen},
    \citenamefont {Stefanucci}, \citenamefont {Almbladh},\ and\
    \citenamefont {von Barth}}]{leeuwen2006c}%
  \BibitemOpen \bibfield {author} {\bibinfo {author} {\bibfnamefont
      {R.}~\bibnamefont {van Leeuwen}}, \bibinfo {author}
    {\bibfnamefont {N.~E.}\ \bibnamefont {Dahlen}}, \bibinfo {author}
    {\bibfnamefont {G.}~\bibnamefont {Stefanucci}}, \bibinfo {author}
    {\bibfnamefont {C.~O.}\ \bibnamefont {Almbladh}}, \ and\ \bibinfo
    {author} {\bibfnamefont {U.}~\bibnamefont {von Barth}},\ }in\
  \href {\doibase 10.1007/3-540-35426-3_3} {\emph {\bibinfo
      {booktitle} {Time-Dependent Density Functional Theory}}},\ Vol.\
  \bibinfo {volume} {706},\ \bibinfo {editor} {edited by\ \bibinfo
    {editor} {\bibfnamefont {M.~A.~L.}\ \bibnamefont {Marques}},
    \bibinfo {editor} {\bibfnamefont {C.~A.}\ \bibnamefont {Ullrich}},
    \bibinfo {editor} {\bibfnamefont {F.}~\bibnamefont {Nogueira}},
    \bibinfo {editor} {\bibfnamefont {A.}~\bibnamefont {Rubio}},
    \bibinfo {editor} {\bibfnamefont {K.}~\bibnamefont {Burke}}, \
    and\ \bibinfo {editor} {\bibfnamefont {E.~K.~U.}\ \bibnamefont
      {Gross}}}\ (\bibinfo {publisher} {Springer},\ \bibinfo {address}
  {Berlin Heidelberg},\ \bibinfo {year}
  {2006})
  \BibitemShut {NoStop}%
\bibitem [{\citenamefont {Rammer}(2007)}]{rammer2007}%
  \BibitemOpen \bibfield {author} {\bibinfo {author} {\bibfnamefont
      {J.}~\bibnamefont {Rammer}},\ }\href@noop {} {\emph {\bibinfo
      {title} {Quantum field theory of nonequilibrium states}}}\
  (\bibinfo {publisher} {Cambridge University Press},\ \bibinfo {year}
  {2007})\BibitemShut {NoStop}%
\bibitem [{}]{convergence}%
  \BibitemOpen Concerning the system size, the equilibrium results are
  converged up to the sixth digit already for 40 sites.  The
  corresponding error is three orders of magnitude smaller than the
  error resulting from the finite time-stepping (see below).
  Consequently, there is in fact no recognizable impact on the
  real-time dynamics when enlarging the system.  Applying the
  trapezoidal rule for all integrations involved on the Keldysh
  branch, we find an error scaling quadratically with the size of the
  time step $\Delta t$ (the error is defined as $\varepsilon(\Delta t)
  = \|V'_{\rm opt}(t;\Delta t) - V'_{\rm opt}(t;\Delta t/2)\|_{\rm
    max}$, i.e., the maximum-norm difference of the optimal parameters
  obtained from calculations with time steps $\Delta t$ and $\Delta
  t/2$). In particular, the error of all our calculations is of the
  order of $\varepsilon \sim 10^{-3}$, i.e., plots for results with a
  smaller time step could hardly be distinguished from the results
  shown. \BibitemShut {NoStop}%
\bibitem [{\citenamefont {Kelley}(1987)}]{kelley1987}%
  \BibitemOpen \bibfield {author} {\bibinfo {author} {\bibfnamefont
      {C.~T.}\ \bibnamefont {Kelley}},\ }\href@noop {} {\emph
    {\bibinfo {title} {Solving nonlinear equations with Newton’s
        method}}},\ Fundamentals of algorithms\ (\bibinfo {publisher}
  {Society for Industrial and Applied Mathematics},\ \bibinfo {year}
  {1987})\BibitemShut {NoStop}%
\bibitem [{\citenamefont {Broyden}(1965)}]{broyden1965}%
  \BibitemOpen \bibfield {author} {\bibinfo {author} {\bibfnamefont
      {C.~G.}\ \bibnamefont {Broyden}},\ }\href {\doibase
    10.1090/S0025-5718-1965-0198670-6} {\bibfield {journal} {\bibinfo
      {journal} {Math. Comp.}\ }\textbf {\bibinfo {volume} {19}},\
    \bibinfo {pages} {577} (\bibinfo {year} {1965})}\BibitemShut
  {NoStop}%
\bibitem [{\citenamefont {Strand}\ \emph {et~al.}(2011)\citenamefont
    {Strand}, \citenamefont {Sabashvili}, \citenamefont {Granath},
    \citenamefont {Hellsing},\ and\ \citenamefont
    {\"Ostlund}}]{strand2011}%
  \BibitemOpen \bibfield {author} {\bibinfo {author} {\bibfnamefont
      {H.~U.~R.}\ \bibnamefont {Strand}}, \bibinfo {author}
    {\bibfnamefont {A.}~\bibnamefont {Sabashvili}}, \bibinfo {author}
    {\bibfnamefont {M.}~\bibnamefont {Granath}}, \bibinfo {author}
    {\bibfnamefont {B.}~\bibnamefont {Hellsing}}, \ and\ \bibinfo
    {author} {\bibfnamefont {S.}~\bibnamefont {\"Ostlund}},\ }\href
  {\doibase 10.1103/PhysRevB.83.205136} {\bibfield {journal} {\bibinfo
      {journal} {Phys. Rev. B}\ }\textbf {\bibinfo {volume} {83}},\
    \bibinfo {pages} {205136} (\bibinfo {year} {2011})}
  \BibitemShut {NoStop}%
\bibitem [{\citenamefont {Bulla}(1999)}]{bulla1999}%
  \BibitemOpen \bibfield {author} {\bibinfo {author} {\bibfnamefont
      {R.}~\bibnamefont {Bulla}},\ }\href {\doibase
    10.1103/PhysRevLett.83.136} {\bibfield {journal} {\bibinfo
      {journal} {Phys. Rev. Lett.}\ }\textbf {\bibinfo {volume}
      {83}},\ \bibinfo {pages} {136} (\bibinfo {year}
    {1999})}\BibitemShut {NoStop}%
\bibitem [{\citenamefont {Lange}(1998)}]{lange1998}%
  \BibitemOpen \bibfield {author} {\bibinfo {author} {\bibfnamefont
      {E.}~\bibnamefont {Lange}},\ }\href {\doibase
    10.1142/S0217984998001050} {\bibfield {journal} {\bibinfo
      {journal} {Modern Physics Letters B}\ }\textbf {\bibinfo
      {volume} {12}},\ \bibinfo {pages} {915} (\bibinfo {year}
    {1998})}
  \BibitemShut {NoStop}%
\bibitem [{\citenamefont {Hubbard}(1963)}]{hubbard1963}%
  \BibitemOpen \bibfield {author} {\bibinfo {author} {\bibfnamefont
      {J.}~\bibnamefont {Hubbard}},\ }\href {\doibase
    10.1098/rspa.1963.0204} {\bibfield {journal} {\bibinfo {journal}
      {Proc. R. Soc. Lond. A}\ }\textbf {\bibinfo {volume} {276}},\
    \bibinfo {pages} {238} (\bibinfo {year} {1963})}\BibitemShut
  {NoStop}%
\bibitem [{\citenamefont {\v{Z}itko}\ and\ \citenamefont
    {Fabrizio}(2015)}]{zitko2015}%
  \BibitemOpen \bibfield {author} {\bibinfo {author} {\bibfnamefont
      {R.}~\bibnamefont {\v{Z}itko}}\ and\ \bibinfo {author}
    {\bibfnamefont {M.}~\bibnamefont {Fabrizio}},\ }\href {\doibase
    10.1103/PhysRevB.91.245130} {\bibfield {journal} {\bibinfo
      {journal} {Phys. Rev. B}\ }\textbf {\bibinfo {volume} {91}},\
    \bibinfo {pages} {245130} (\bibinfo {year} {2015})}
  \BibitemShut {NoStop}%
\bibitem [{\citenamefont {de'Medici}\ \emph
    {et~al.}(2005)\citenamefont {de'Medici}, \citenamefont {Georges},\
    and\ \citenamefont {Biermann}}]{medici2005b}%
  \BibitemOpen \bibfield {author} {\bibinfo {author} {\bibfnamefont
      {L.}~\bibnamefont {de'Medici}}, \bibinfo {author} {\bibfnamefont
      {A.}~\bibnamefont {Georges}}, \ and\ \bibinfo {author}
    {\bibfnamefont {S.}~\bibnamefont {Biermann}},\ }\href {\doibase
    10.1103/PhysRevB.72.205124} {\bibfield {journal} {\bibinfo
      {journal} {Phys. Rev. B}\ }\textbf {\bibinfo {volume} {72}},\
    \bibinfo {pages} {205124} (\bibinfo {year} {2005})}
  \BibitemShut {NoStop}%
\bibitem [{\citenamefont {R\"uegg}\ \emph {et~al.}(2010)\citenamefont
    {R\"uegg}, \citenamefont {Huber},\ and\ \citenamefont
    {Sigrist}}]{rueegg2010}%
  \BibitemOpen \bibfield {author} {\bibinfo {author} {\bibfnamefont
      {A.}~\bibnamefont {R\"uegg}}, \bibinfo {author} {\bibfnamefont
      {S.~D.}\ \bibnamefont {Huber}}, \ and\ \bibinfo {author}
    {\bibfnamefont {M.}~\bibnamefont {Sigrist}},\ }\href {\doibase
    10.1103/PhysRevB.81.155118} {\bibfield {journal} {\bibinfo
      {journal} {Phys. Rev. B}\ }\textbf {\bibinfo {volume} {81}},\
    \bibinfo {pages} {155118} (\bibinfo {year} {2010})}
  \BibitemShut {NoStop}%
\bibitem [{\citenamefont {Francuz}\ \emph {et~al.}(2016)\citenamefont
    {Francuz}, \citenamefont {Dziarmaga}, \citenamefont {Gardas},\
    and\ \citenamefont {Zurek}}]{francuz2016}%
  \BibitemOpen \bibfield {author} {\bibinfo {author} {\bibfnamefont
      {A.}~\bibnamefont {Francuz}}, \bibinfo {author} {\bibfnamefont
      {J.}~\bibnamefont {Dziarmaga}}, \bibinfo {author} {\bibfnamefont
      {B.}~\bibnamefont {Gardas}}, \ and\ \bibinfo {author}
    {\bibfnamefont {W.~H.}\ \bibnamefont {Zurek}},\ }\href {\doibase
    10.1103/PhysRevB.93.075134} {\bibfield {journal} {\bibinfo
      {journal} {Phys.  Rev. B}\ }\textbf {\bibinfo {volume} {93}},\
    \bibinfo {pages} {075134} (\bibinfo {year}
    {2016})}
  \BibitemShut {NoStop}%
\end{thebibliography}


%

\end{document}